\documentclass[a4paper,useAMS,usenatbib]{mn2e}
\usepackage{amsmath}
\usepackage{amssymb}
\usepackage{graphicx}
\usepackage{esint}
\usepackage[authoryear]{natbib}

 \PassOptionsToPackage{caption=false}{subfig}
\usepackage{subfig}

\makeatletter

\pdfpageheight\paperheight
\pdfpagewidth\paperwidth

\title[Generalized Formalisms of the RIME]{Generalized Formalisms of the Radio Interferometer Measurement Equation}

\author[D.C.~Price and O.~Smirnov]{
D. C. Price$^{1}$\thanks{E-mail: dprice@cfa.harvard.edu}
and O. M. Smirnov$^{2,3}$\\
$^1$Harvard-Smithsonian Center for Astrophysics, MS42, 60 Garden Street, Cambridge MA, 01243 United States\\
$^2$Department of Physics and Electronics, Rhodes University, PO Box 94, Grahamstown, 6140, South Africa\\
$^3$SKA South Africa, 3rd Floor, The Park, Park Road, Pinelands, 7405, South Africa
\\}

\makeatother

\begin{document}

\date{Accepted 2015 January 12. Received 2014 November 04}

\maketitle
\pagerange{\pageref{firstpage}--\pageref{lastpage}} \pubyear{2015}

\label{firstpage}
\begin{abstract}
The Radio Interferometer Measurement Equation (RIME) is a matrix-based mathematical
model that describes the response of a radio interferometer.  The Jones
calculus it employs is not suitable for describing the analogue components
of a telescope. This is because it does not consider the effect of impedance
mismatches between components.  This paper aims to highlight the limitations
of Jones calculus, and suggests some alternative methods that are
more applicable.
We reformulate the RIME with a different basis that includes magnetic and mixed coherency statistics. 
We present a microwave network inspired 2N-port version of the RIME, and a tensor formalism based upon
the electromagnetic tensor from special relativity.
We elucidate the limitations of the Jones-matrix-based RIME for describing analogue components. We show how
measured scattering parameters of analogue components can be used in a 2N-port version of the RIME. In addition, we 
show how motion at relativistic speed affects the observed flux.  
We present reformulations of the RIME that correctly account for magnetic field coherency. 
These reformulations extend the standard formulation, highlight its limitations, and may have applications in
space-based interferometry and precise absolute calibration experiments.
   
\end{abstract}


\section{Introduction}

Coherency in electromagnetic fields is a fundamental topic within
optics. Its importance in fields such as radio astronomy can not be
overstated: interferometry and synthesis imaging techniques rely heavily
upon coherency theory \citep{Taylor1999,BookMandelWolf,ThompsonMoranSwenson2004}.
Of particular importance to radio astronomy is the Van-Cittert-Zernicke
theorem (vC-Z, \citealp{VanCittertZernicke1938}) and the radio interferometer
Measurement Equation (RIME, \citealp{Hamaker:1996p5735}). The vC-Z
relates the brightness of a source to its mutual coherency as measured
by an interferometer, and the RIME provides a polarimetric framework
to calibrate out corruptions caused along the signal's path. 

While the vC-Z theorem dates back to 1938, more recent work such as
that of \citet{Carozzi:2009hf} extends its applicability to polarized
measurements over wide fields of view. The RIME has a much shorter
history: it was not formulated until 1996 \citep{Hamaker:1996p5735}.
Before the RIME, calibration was conducted in an ad-hoc manner, with
each polarization essentially treated separately. The framework was
expounded in a series of follow-up papers \citep{Sault:1996p5731,Hamaker:1996p5733,Hamaker:2000p7625,Hamaker:2006p7626}; recent work by Smirnov extends the formalism to the full sky case,
and reformulates the RIME using tensor algebra \citep{Smirnov:2011a,Smirnov:2011d}. 

This article introduces two reformulations of the RIME, that extend its applicability and demonstrates
limitations with the Jones-matrix-based formalism. 
Both of these reformulations consider full electromagnetic coherency statistics
(i.e. electric, magnetic and mixed coherencies). The first is inspired by transmission matrix methods
from microwave engineering. We show that this reformulation better accounts for the changes in 
impedance between analogue components within a radio telescope.
The second reformulation is a relativistic-aware formulation of the RIME,
starting with the electromagnetic field tensor. This formalism allows for relativistic motion to be treated
as instrumental effect and incorporated into the RIME.

This article is organized as follows. In Section 2, we review existing formalisms of the RIME and methods from 
microwave engineering. Section 3 defines coherency matrices, which are used in Section 4 to formulate 
general coherency relationships for radio astronomy. In Section 5, we introduce a tensor formulation of the RIME based upon
the electromagnetic tensor from special relativity. Discussion and example applications are given in Section 6;
concluding remarks are given in Section 7.

\section{Jones and Mueller RIME formulations}

Before continuing on to derive a more general relationship between
the two-point coherency matrix and the voltage-current coherency,
we would like to give a brief overview and derivation of the radio
interferometer Measurement Equation of \citet{Hamaker:1996p5735}.
Our motivation behind this is to highlight that Hamaker et. al.'s
RIME is a special case of a more general (and thus less limited) coherency
relationship.

In their seminal Measurement Equation (ME) paper, \citet{Hamaker:1996p5735}
showed that Mueller and Jones calculuses provide a good framework for
modelling radio interferometers. In optics, Jones and Mueller matrices
are used to model the transmission of light through optical elements
\citep{Jones1941,Mueller1948}. Mueller matrices are $\mbox{4}\times\mbox{4}$
matrices that act upon the Stokes vector 
\begin{equation}
\mbox{\textbf{\emph{s}}}=\begin{pmatrix}I & Q & U & V\end{pmatrix}^{T},
\end{equation}
whereas Jones matrices are only $\mbox{2}\times\mbox{2}$ in dimension
and act upon the `Jones vector': the electric field vector in a coordinate
system such that z-axis is aligned with the Poynting vector 
\begin{equation}
\mbox{\textbf{\emph{e}}}(\mbox{\emph{\textbf{r}}},t)=\begin{pmatrix}e_{x}(\mbox{\emph{\textbf{r}}},t) & e_{y}(\mbox{\emph{\textbf{r}}},t)\end{pmatrix}^{T}.
\end{equation}
Jones calculus dictates that along a signal's path, any (linear) transformation
can be represented with a Jones matrix, \textbf{\emph{J}}:
\begin{equation}
\mbox{\textbf{\emph{e}}}_{\rm{out}}(\mbox{\textbf{\emph{r}}},t)=\mbox{\textbf{\emph{J}}}\mbox{\textbf{\emph{e}}}_{\rm{in}}(\mbox{\textbf{\emph{r}}},t)\label{eq:jones-transmission}
\end{equation}
A useful property of Jones calculus is that multiple effects along
a signal's path of propagation correspond to repeated multiplications:
\begin{equation}
\mbox{\textbf{\emph{e}}}_{\mathrm{out}}(\mbox{\textbf{\emph{r}}},t)=\emph{\textbf{J}}_{n}\cdots\emph{\textbf{J}}_{2}\emph{\textbf{J}}_{1}\mbox{\textbf{\emph{e}}}_{\rm{in}}(\mbox{\textbf{\emph{r}}},t),
\end{equation}
which can be collapsed into a single matrix when required.

\begin{figure}
\begin{centering}
\includegraphics[width=0.95\columnwidth]{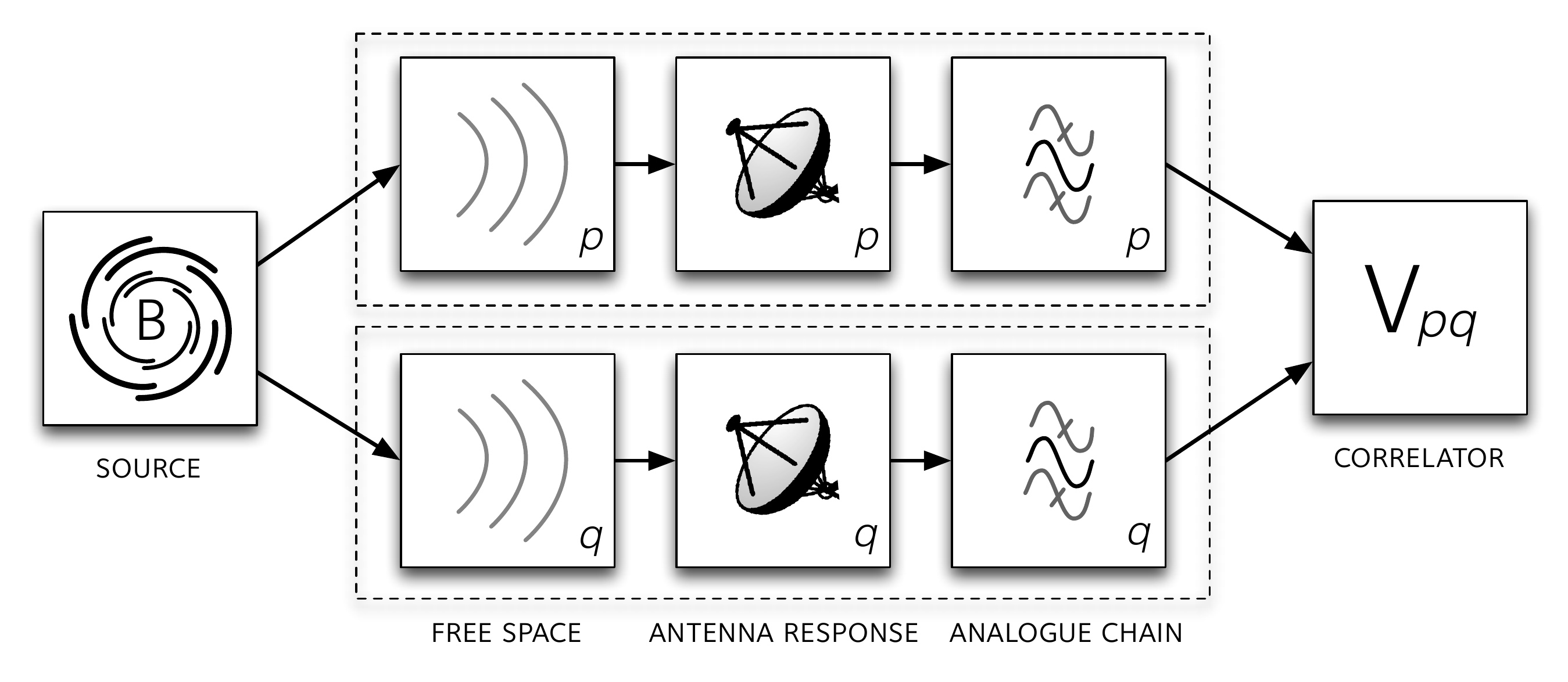}
\par\end{centering}

\caption{Block diagram showing a simple model of an interferometer that can
be modelled with the RIME. Radiation from a source propagates through
free space to two telescopes, \emph{p }and \emph{q}. After passing
through the telescope's analogue chain, the two signals are interfered
in a cross-correlator.\label{fig:RIME-cartoon}}
\end{figure}

The RIME uses Jones matrices to model the various corruptions and
effects during a signal's journey from a source right though to the
correlator. A block diagram for a (simplified) two-element interferometer
is shown in Figure~\ref{fig:RIME-cartoon}. From left to right, the
figure shows the journey of a signal from a source right through to
the correlator. The radiation from the source is picked up by two
antennas, which we have denoted with subscript $p$ and $q$. The
radiation follows a unique path to both of these antennas; each antenna
also has associated with it a unique chain of analogue components
that amplify and filter the signal to prepare it for correlation.
Each of these subscripted boxes may be represented by a Jones matrix;
alternatively an overall Jones matrix can be formed for the $p$ and
$q$ branches (the dashed areas).

\subsection{Hamaker's RIME derivation}

The derivation of the RIME is remarkably simple and elegant. For a
single point source of radiation, the voltages induced at the terminals
of a pair of antennas, $p$ and $q$ are 
\begin{align}
\mbox{\textbf{\emph{v}}}_{p}(t) & =\mbox{\textbf{\emph{J}}}_{p}\mbox{\textbf{\emph{e}}}_{0}(t)\\
\mbox{\textbf{\emph{v}}}_{q}(t) & =\mbox{\textbf{\emph{J}}}_{q}\mbox{\textbf{\emph{e}}}_{0}(t)
\end{align}
In its simplest form, the RIME is formed by taking the outer product
of these two relationships. Note that in their original paper, the
authors use the Kronecker incarnation of the outer product, which
we will denote with $\star$. We reserve the symbol $\otimes$ for
the matrix outer product of two matrices $\emph{\textbf{A}}\otimes\emph{\textbf{B}}=\mbox{\textbf{\emph{A}}}\emph{\textbf{B}}^{H},$
where \emph{H} denotes the Hermitian conjugate transpose%
\footnote{For a discussion on the subtleties of outer product definition, see
\citealt{Smirnov:2011d}, §A.6.1%
}. Using the Kronecker outer product, the RIME is given by
\begin{equation}
\left\langle \mbox{\textbf{\emph{v}}}_{p}\star\mbox{\textbf{\emph{v}}}_{q}\right\rangle =\left(\mbox{\textbf{\emph{J}}}_{p}\star\mbox{\textbf{\emph{J}}}_{q}\right)\left\langle \mbox{\textbf{\emph{e}}}_{0}\star\mbox{\textbf{\emph{e}}}_{0}\right\rangle =\left(\mbox{\textbf{\emph{J}}}_{p}\star\mbox{\textbf{\emph{J}}}_{q}\right)\mbox{\textbf{\emph{e}}}_{00}\label{eq:mueller-rime}
\end{equation}
where $\mbox{\textbf{\emph{J}}}_{p}$ and $\mbox{\textbf{\emph{J}}}_{q}$
are the Jones matrices representing all transformations along the
two signal paths, and $\left(\mbox{\textbf{\emph{J}}}_{p}\star\mbox{\textbf{\emph{J}}}_{q}\right)$
is a $\mbox{4}\times\mbox{4}$ matrix. Here, $\mbox{\textbf{\emph{e}}}_{00}$
is the sky brightness of a single point source of radiation. For
a multi-element interferometer, every antenna has its own unique Jones
matrix, and a RIME may be written for every pair of antennas.

Due to their choice of outer product, Hamaker et. al. arrive at a
coherency vector
\begin{equation}
\mbox{\textbf{\emph{e}}}_{pq}(\mbox{\textbf{\emph{r}}}_{p},\mbox{\textbf{\emph{r}}}_{q},\tau)=\left\langle \mbox{\textbf{\emph{e}}}_{p}(\mbox{\textbf{\emph{r}}}_{p},t)\star\mbox{\textbf{\emph{e}}}_{q}(\mbox{\textbf{\emph{r}}}_{q},t+\tau)\right\rangle =\begin{pmatrix}\left\langle e_{px}e_{qx}^{*}\right\rangle \\
\left\langle e_{px}e_{qy}^{*}\right\rangle \\
\left\langle e_{py}e_{qx}^{*}\right\rangle \\
\left\langle e_{py}e_{qy}^{*}\right\rangle 
\end{pmatrix},
\end{equation}
as opposed to the coherency matrix of \citet{Wolf1954}; this is introduced
in §\ref{sub:Electromagnetic-coherency} below. The column vector
of a point source at $\mbox{\textbf{\emph{r}}}_{0}$ is then $\mbox{\textbf{\emph{e}}}_{00}$;
that is, $p=q$ and $\tau=0$. The vector $\mbox{\textbf{\emph{e}}}_{00}$
is related to the Stokes vector by the transform\footnote{Here, $j=\sqrt{-1}$, to avoid confusion with current, $i$, used later.}.
\begin{equation}
\begin{pmatrix}I\\
Q\\
U\\
V
\end{pmatrix}=\begin{pmatrix}1 & 0 & 0 & 1\\
1 & 0 & 0 & -1\\
0 & 1 & 1 & 0\\
0 & -j & j & 0
\end{pmatrix}\begin{pmatrix}\left\langle e_{0x}e_{0x}^{*}\right\rangle \\
\left\langle e_{0x}e_{0y}^{*}\right\rangle \\
\left\langle e_{0y}e_{0x}^{*}\right\rangle \\
\left\langle e_{0y}e_{0y}^{*}\right\rangle 
\end{pmatrix}.
\end{equation}
The quantity $\left(\mbox{\textbf{\emph{J}}}_{p}\star\mbox{\textbf{\emph{J}}}_{q}\right)$
in Eq.~\ref{eq:mueller-rime} can therefore be viewed as a Mueller
matrix. That is, Eq.~\ref{eq:mueller-rime} can be considered a Mueller-calculus-based ME for a radio interferometer. To
summarize, Hamaker et. al. showed that:
\begin{itemize}
\item Jones matrices can be used to model the propagation of a signal from
a radiation source through to the voltage at the terminal of an antenna.
\item A Mueller matrix can be formed from the Jones terms of a pair of antennas,
which then relates the measured voltage coherency of that pair to
a source's brightness.
\end{itemize}
Showing that these calculuses were applicable and indeed useful for modelling
and calibrating radio interferometers was an important step forward
in radio polarimetry.

\subsection{The 2$\times$2 RIME}

In a later paper, \citet{Hamaker:2000p7625} presents a modified formulation
of the RIME, where instead of forming the coherency vector from the Kronecker
outer product ($\star$), the coherency matrix is formed from the matrix
outer product ($\otimes$):
\begin{equation}
\mbox{\textbf{\emph{E}}}_{pq}=\left\langle \emph{\textbf{e}}_{p}\otimes\emph{\textbf{e}}_{q}\right\rangle =\begin{pmatrix}\left\langle e_{px}e_{qx}^{*}\right\rangle  & \left\langle e_{px}e_{qy}^{*}\right\rangle \\
\left\langle e_{px}e_{qx}^{*}\right\rangle  & \left\langle e_{py}e_{qy}^{*}\right\rangle 
\end{pmatrix}
\end{equation}
 The resulting coherency matrix is then shown to be related to the
Stokes parameters by $\mbox{\textbf{\emph{E}}}_{00}=\mbox{ \textsf{B} }$,
where 
\begin{align}
\mathsf{B} & =\begin{pmatrix}I+Q & U+jV\\
U-jV & I-Q
\end{pmatrix}.
\end{align}
The equivalent to the RIME of Eq.~\ref{eq:mueller-rime} is 
\begin{equation}
\left\langle \mbox{\textbf{\emph{v}}}_{p}\otimes\mbox{\textbf{\emph{v}}}_{q}\right\rangle =2\left\langle \mbox{(\ensuremath{\mbox{\textbf{\emph{J}}}_{p}\textbf{\emph{e}}}}_{0})\otimes(\mbox{\textbf{\emph{J}}}_{q}\mbox{\textbf{\emph{e}}}_{0})\right\rangle =2\mbox{\textbf{\emph{J}}}_{p}\left\langle \mbox{\textbf{\emph{e}}}_{0}\otimes\mbox{\textbf{\emph{e}}}_{0}\right\rangle \mbox{\textbf{\emph{J}}}_{q}^{H}\label{eq:jones-rime}
\end{equation}
or more simply, 
\begin{equation}
\mathsf{V}_{pq}=\mbox{\textbf{\emph{J}}}_{p}\mbox{\textbf{\emph{B}}}\mbox{\textbf{\emph{J}}}_{q}^{H}.\label{eq:jones-rime-2}
\end{equation}
This approach avoids the need to use $\mbox{4}\times\mbox{4}$ Mueller
matrices, so is both simpler and computationally advantageous. This
form is also cleaner in appearance, as fewer indices are required.

\citet{Smirnov:2011a} takes the $\mbox{2}\times\mbox{2}$ version
of the RIME as a starting point and extends the RIME to a full sky
case. By treating the sky as a brightness distribution $\mathsf{B}$($\sigma$),
where $\sigma$ is a direction vector, each antenna has a Jones term
$\mbox{\textbf{\emph{J}}}_{p}(\sigma)$ describing the signal path
for a given direction. The visibility matrix $\mathsf{V}_{pq}$ is
then found by integrating over the entire sky:
\begin{equation}
\mathsf{V}_{pq}=\int_{4\pi}\mbox{\textbf{\emph{J}}}_{p}(\sigma)\mathsf{B}\mbox{(\ensuremath{\sigma})}\mbox{\textbf{\emph{J}}}_{q}^{H}(\sigma)d\Omega.
\end{equation}
This is a more general form of\emph{ Zernicke's propagation law}.
Smirnov goes on to derive the Van-Cittert Zernicke theorem from this
result; we return to vC-Z later in this article.

\subsection{A generalized tensor RIME}

  In \citet{Smirnov:2011d}, a generalized tensor formalism of the
RIME is presented. The coherency of two voltages is once again defined
via the outer product $e^{i}\bar{e}_{j}$,  giving a (1,1)-type
tensor expression:
\begin{equation}
\mbox{\emph{V}}_{qj}^{pi}=\mbox{\emph{J}}_{\alpha}^{pi}\mbox{\emph{B}}_{\beta}^{\alpha}\bar{\mbox{\emph{J}}}_{qj}^{\beta}.
\end{equation}
This formalism is better capable of describing mutual coupling between
antennas, beamforming, and wide field polarimetry. $ $ In this paper,
we focus on a matrix based formalism which considers the propagation
of the magnetic field in addition to the electric field. We then show
that this formulation is equivalent to the tensor formalism presented
in \citet{Smirnov:2011d}, but is instead in the vector space $\mathbb{C}^{6}$.

\subsection{Microwave engineering transmission matrix methods}

All formulations of the RIME to date --- including the tensor formulation
--- do not consider the propagation of the magnetic field. In free
space, magnetic field coherency can be easily derived from the electric
field coherency. However, at the boundary between two media, the magnetic
field must be considered. Here, we introduce some results from microwave
engineering which contrast with the Jones formalism.

In circuit theory, the well-known impedance relation, $V=ZI$ relates
current and voltage over a terminal pair (or `port'). However, this
relation is specific to a 1-port network; for microwave networks with
more than one port, the matrix form $[V]=[Z][I]$ must be used:

\begin{equation}
\begin{pmatrix}v_{1}\\
v_{2}\\
\vdots\\
v_{N}
\end{pmatrix}=\begin{pmatrix}Z_{11} & Z_{12} & \cdots & Z_{1N}\\
Z_{21} & \ddots &  & \vdots\\
\vdots &  &  & \vdots\\
Z_{N1} & \cdots & \cdots & Z_{NN}
\end{pmatrix}\begin{pmatrix}i_{1}\\
i_{2}\\
\vdots\\
i_{N}
\end{pmatrix},
\end{equation}
where $Z_{ab}$ is the port-to-port impedance from port \emph{a }to
port \emph{b}, $v_{n}$ is the voltage on port $n$, and $i_{n}$
is the current. A common example of a 2-port network is a coaxial cable, and a common
3-port network is the Wilkinson power divider. 

\begin{figure}
\begin{centering}
\includegraphics[width=0.8\columnwidth]{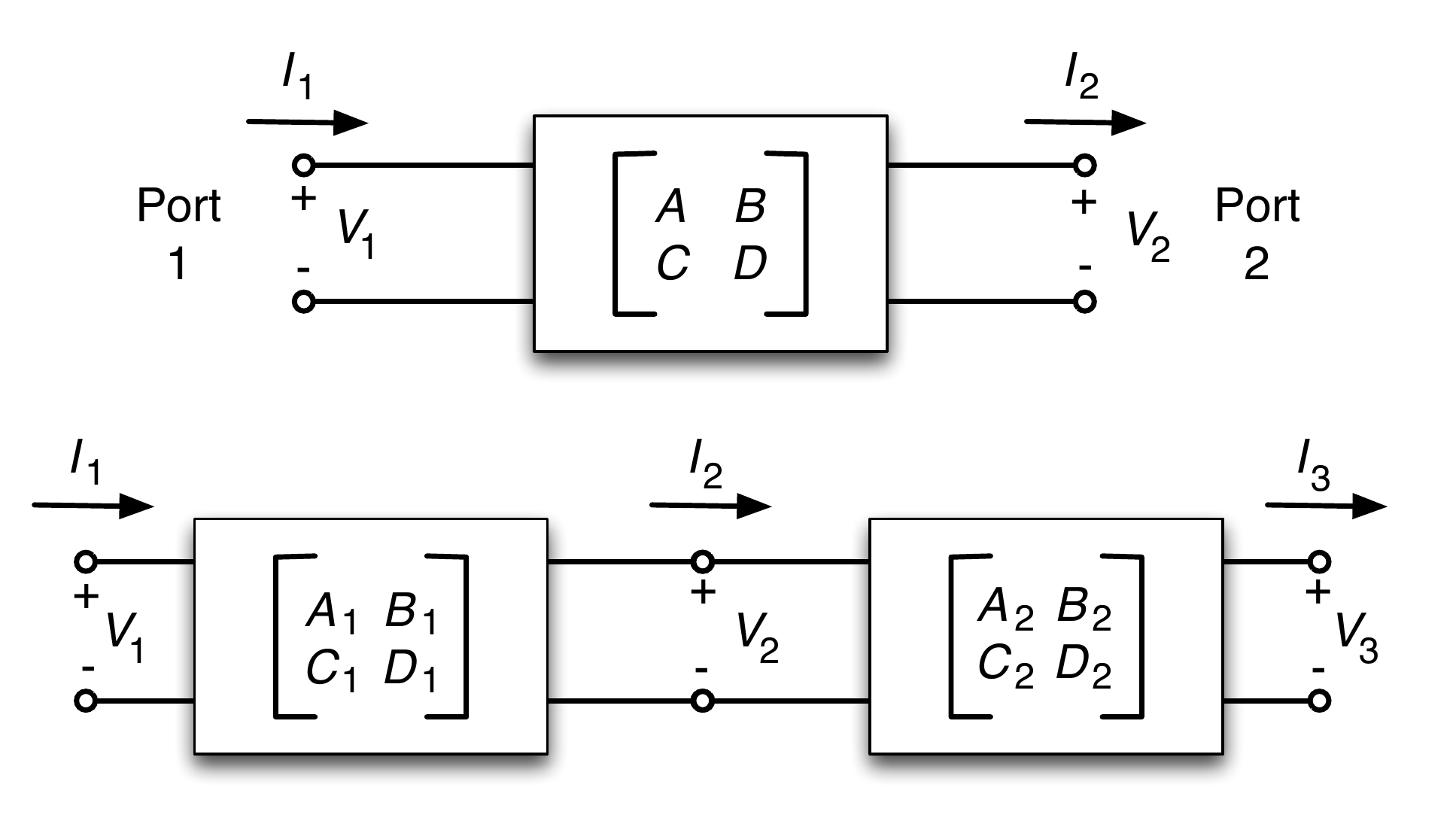}
\par\end{centering}

\caption{Top: The ABCD matrix for a 2-port network. In this diagram, voltage
is denoted with $V$, and current with $I$. Bottom: Connecting two
components in cascade. Diagram adapted from \citet{Pozar2005}\label{fig:transmission-cascade}}
\end{figure}

The analogue components of most radio telescopes can be considered
2-port networks. The 2-port transmission, or $ABCD$ matrix, relates
the voltages and currents of a 2-port network:

\begin{equation}
\begin{pmatrix}v_{1}\\
i_{1}
\end{pmatrix}=\begin{pmatrix}A & B\\
C & D
\end{pmatrix}\begin{pmatrix}v_{2}\\
i_{2}
\end{pmatrix},
\end{equation}
this is shown in Figure~\ref{fig:transmission-cascade}. If two 2-port
networks are connected in cascade (see Figure~\ref{fig:transmission-cascade}),
then the output is equal to the product of the transmission matrices
representing the individual components:
\begin{equation}
\begin{pmatrix}v_{1}\\
i_{1}
\end{pmatrix}=\begin{pmatrix}A_{1} & B_{1}\\
C_{1} & D_{1}
\end{pmatrix}\begin{pmatrix}A_{2} & B_{2}\\
C_{2} & D_{2}
\end{pmatrix}\begin{pmatrix}v_{2}\\
i_{2}
\end{pmatrix},
\end{equation}
as is shown in texts such as \citet{Pozar2005}. The elements in the
2-port transmission matrix are related to port-to-port impedances
by%
\footnote{Note that $Z_{21}=0$ is zero impedance, which is never satisfied
in real components, and $Z_{21}=\infty$ represents an open circuit%
} 
\begin{align}
A & =Z_{11}/Z_{21}\\
B & =\frac{Z_{11}Z_{22}-Z_{12}Z_{21}}{Z_{21}}\\
C & =1/Z_{21}\\
D & =Z_{22}/Z_{21}.
\end{align}
Like the Jones matrix, the $ABCD$ matrix allows a signal's path to
be modelled through multiplication of matrices representing discrete
components. While the Jones matrix acts upon a pair of orthogonal
electric field components, the $ABCD$ matrix acts upon a voltage-current	
pair at a single port. As Jones matrices do not consider changes in
impedance (free space impedance is implicitly assumed), it is not
suitable for describing analogue components. Conversely, the $2\times2$
$ABCD$ matrix cannot model cross-polarization response of a telescope.
In the section that follows, we derive a more general coherency relationship
which weds the advantages of both approaches.

\section{Coherency in radio astronomy}

We now turn our attention to formulating a more general RIME that
is valid in a larger range of cases. In this section, we introduce
the coherency matrices of \citet{Wolf1954}, along with voltage-current
coherency matrices. The following section then formulates relationships
between source brightness and measured coherency based upon these
matrices.

\subsection{Electromagnetic coherency\label{sub:Electromagnetic-coherency}}

To begin, we introduce the coherency matrices of \citet{Wolf1954},
that fully describe the coherency statistics of an electromagnetic
field. We may start by introducing $\mbox{\textbf{\emph{e}}}(\mbox{\textbf{\emph{r}}},t)$
and $\mbox{\textbf{\emph{h}}}(\mbox{\textbf{\emph{r}}},t)$ as the
complex analytic representations of the electric and magnetic field
vectors at a spacetime point $(\mbox{\textbf{\emph{r}}},t)$: 
\begin{align}
\mbox{\textbf{\emph{e}}}(\mbox{\textbf{\emph{r}}},t) & =\begin{pmatrix}e_{x}(\mbox{\textbf{\emph{r}}},t) & e_{y}(\mbox{\textbf{\emph{r}}},t) & e_{z}(\mbox{\textbf{\emph{r}}},t)\end{pmatrix}^{T}\label{eq:elec-vec}\\
\mbox{\textbf{\emph{h}}}(\mbox{\textbf{\emph{r}}},t) & =\begin{pmatrix}h_{x}(\mbox{\textbf{\emph{r}}},t) & h_{y}(\mbox{\textbf{\emph{r}}},t) & h_{z}(\mbox{\textbf{\emph{r}}},t)\end{pmatrix}^{T}
\end{align}
The coherency matrices are then defined by the formulae
\begin{align}
\mbox{\textbf{\emph{E}}}_{pq}(\mbox{\textbf{\emph{r}}}_{p},\mbox{\textbf{\emph{r}}}_{q},\tau) & =\begin{pmatrix}\left\langle e_{k}(\mbox{\textbf{\emph{r}}}_{p},t)e_{l}^{*}(\mbox{\textbf{\emph{r}}}_{q},t+\tau)\right\rangle \end{pmatrix}\\
\mbox{\textbf{\emph{H}}}_{pq}(\mbox{\textbf{\emph{r}}}_{p},\mbox{\textbf{\emph{r}}}_{q},\tau) & =\begin{pmatrix}\left\langle h_{k}(\mbox{\textbf{\emph{r}}}_{p},t)h_{l}^{*}(\mbox{\textbf{\emph{r}}}_{q},t+\tau)\right\rangle \end{pmatrix}\\
\mbox{\textbf{\emph{M}}}_{pq}(\mbox{\textbf{\emph{r}}}_{p},\mbox{\textbf{\emph{r}}}_{q},\tau) & =\begin{pmatrix}\left\langle e_{k}(\mbox{\textbf{\emph{r}}}_{p},t)h_{l}^{*}(\mbox{\textbf{\emph{r}}}_{q},t+\tau)\right\rangle \end{pmatrix}\\
\mbox{\textbf{\emph{N}}}_{pq}(\mbox{\textbf{\emph{r}}}_{p},\mbox{\textbf{\emph{r}}}_{q},\tau) & =\begin{pmatrix}\left\langle h_{k}(\mbox{\textbf{\emph{r}}}_{p},t)e_{l}^{*}(\mbox{\textbf{\emph{r}}}_{q},t+\tau)\right\rangle \end{pmatrix}.
\end{align}
Here, $k$ and $l$ are indices representing the $x,y,z$ subscripts
from Cartesian coordinates. $\mbox{\textbf{\emph{E}}}_{pq}$ and $\mbox{\textbf{\emph{H}}}_{pq}$
are called the electric and the magnetic coherency matrices, and $\mbox{\textbf{\emph{M}}}_{pq}$
and $\mbox{\textbf{\emph{N}}}_{pq}$ are called the mixed coherency
matrices. The subscripts $p$ and $q$ correspond to the spacetime
points $(\mbox{\textbf{\emph{r}}}_{p},t)$ and $(\mbox{\textbf{\emph{r}}}_{q},t+\tau)$,
respectively. We may arrange these into a single $\mbox{6}\times\mbox{6}$
matrix \textbf{$\mathbb{B}_{pq}$} that is equivalent to the time
averaged outer product of the electromagnetic field column vectors
at spacetime points $(\mbox{\textbf{\emph{r}}}_{p},t)$ and $(\mbox{\textbf{\emph{r}}}_{q},t+\tau)$:
\begin{equation}
\mathbb{B}_{pq}=\left\langle \begin{pmatrix}\mbox{\textbf{\emph{e}}}_{p}\\
\mbox{\textbf{\emph{h}}}_{p}
\end{pmatrix}\otimes\begin{pmatrix}\mbox{\textbf{\emph{e}}}_{q}\\
\mbox{\textbf{\emph{h}}}_{q}
\end{pmatrix}\right\rangle =\begin{pmatrix}\mbox{\textbf{\emph{E}}}_{pq} & \mbox{\textbf{\emph{M}}}_{pq}\\
\mbox{\textbf{\emph{N}}}_{pq} & \mbox{\textbf{\emph{H}}}_{pq}
\end{pmatrix}\label{eq:coherency-matrix}
\end{equation}
This matrix fully describes the coherency properties of an electromagnetic
field at two points in spacetime. We will refer to this matrix as
the \emph{two point coherency matrix}. It is worth noting that:
\begin{itemize}
\item When $\mbox{\textbf{\emph{r}}}_{p}=\mbox{\textbf{\emph{r}}}_{q}$
and $\tau=0$ we retrieve what \citet{Bergman:2008p7859} refer to
as the `EM sixtor matrix'. \citet{Bergman:2008p7859} show this sixtor
matrix is related to what they refer to as `canonical electromagnetic
observables': a unique set of Stokes-like parameters that are irreducible
under Lorentz transformations. These are used in the the analysis
of electromagnetic field data from spacecraft.
\item For monochromatic plane waves, when $\mbox{\textbf{\emph{r}}}_{p}=\mbox{\textbf{\emph{r}}}_{q}$
and $\tau=0$, and we choose a coordinate system with $z$ in the
direction of propagation (i.e. along the Poynting vector), $\textbf{\emph{E}}_{pq}$ 
becomes what \citet{Smirnov:2011a} refers to as the \emph{brightness
matrix}, $\mathsf{B}$. 

From here forward, we drop the subscript $\mathbb{B}=\mathbb{B}_{00}$ and shall refer to 
this as the \emph{brightness coherency} to highlight its relationship with $\mathsf{B}$.
\end{itemize}

\subsection{Voltage and current coherency}

A radio telescope converts a free space electromagnetic field into
a time varying voltage, which we then measure after signal conditioning
(e.g. amplification and filtering). As such, radio interferometers
measure coherency statistics between time varying voltages. 

One may model the analogue components of a telescope as a 6-port network,
with three inputs ports and three output ports. We propose this so
that there is an input-output pair of ports for each of the orthogonal
components of the electromagnetic field. We can then define a set
of voltages \textbf{\emph{v}}\emph{(t)} and currents $\mbox{\textbf{\emph{i}}}(t)$

\begin{align}
\mbox{\textbf{\emph{v}}}(t) & =\begin{pmatrix}v_{x}(t) & v_{y}(t) & v_{z}(t)\end{pmatrix}^{T}\\
\mbox{\textbf{\emph{i}}}(t) & =\begin{pmatrix}i_{x}(t) & i_{y}(t) & i_{z}(t)\end{pmatrix}^{T}.
\end{align}
In practice, most telescopes are single or dual polarization, so only
the $x$ and $y$ components are sampled. Nonetheless, it is possible
to sample all three components with three orthogonal antenna elements
\citep{Bergman:2005p7825}. The voltage response of an antenna is
linearly related to the electromagnetic field strength \citep{Hamaker:1996p5735},
and the current is linearly related to voltage by Ohm's law, so we
may write a general linear relationship 
\begin{equation}
\begin{pmatrix}\mbox{\textbf{\emph{v}}}(t)\\
\mbox{\textbf{\emph{i}}}(t)
\end{pmatrix}=\begin{pmatrix}\mbox{\textbf{\emph{A}}} & \mbox{\textbf{\emph{B}}}\\
\mbox{\textbf{\emph{C}}} & \mathbf{\mbox{\textbf{\emph{D}}}}
\end{pmatrix}\begin{pmatrix}\mbox{\textbf{\emph{e}}}(\mbox{\textbf{\emph{r}}},t)\\
\mbox{\textbf{\emph{h}}}(\mbox{\textbf{\emph{r}}},t)
\end{pmatrix},
\end{equation}
where \textbf{\emph{A}}, \textbf{\emph{B}},\textbf{ }\textbf{\emph{C}}\textbf{
}and\textbf{ }\textbf{\emph{D}}\textbf{ }are block matrices forming
an overall transmission matrix $\mathbb{T}'$. We can now define a
matrix of voltage-current coherency statistics that consists of the
block matrices

\begin{align}
\mbox{\textbf{\emph{V}}}_{pq}(\tau) & =\begin{pmatrix}\left\langle v_{k}(t)v_{l}^{*}(t+\tau)\right\rangle \end{pmatrix}\\
\mbox{\textbf{\emph{W}}}_{pq}(\tau) & =\begin{pmatrix}\left\langle i_{k}(t)i_{l}^{*}(t+\tau)\right\rangle \end{pmatrix}\\
\mbox{\textbf{\emph{K}}}_{pq}(\tau) & =\begin{pmatrix}\left\langle v_{k}(t)i_{l}^{*}(t+\tau)\right\rangle \end{pmatrix}\\
\mbox{\textbf{\emph{L}}}_{pq}(\tau) & =\begin{pmatrix}\left\langle i_{k}(t)v_{l}^{*}(t+\tau)\right\rangle \end{pmatrix},
\end{align}
these are analogous to (and related to)\textbf{ }the electromagnetic
coherency matrices above%
\footnote{Spatial location $\mathbf{r}$ is no longer relevant as the voltage
propagates through analogue components with clearly defined inputs
and outputs.%
}. In a similar manner to the two-point coherency matrix, we define
$\mathbb{V}_{pq}$ 
\begin{equation}
\mathbb{V}_{pq}=\left\langle \begin{pmatrix}\mbox{\textbf{\emph{v}}}_{p}\\
\mbox{\textbf{\emph{i}}}_{p}
\end{pmatrix}\otimes\begin{pmatrix}\mbox{\textbf{\emph{v}}}_{q}\\
\mbox{\textbf{\emph{i}}}_{q}
\end{pmatrix}\right\rangle =\begin{pmatrix}\mbox{\textbf{\emph{V}}}_{pq} & \mathbf{\mbox{\textbf{\emph{K}}}}_{pq}\\
\mbox{\textbf{\emph{L}}}_{pq} & \mbox{\textbf{\emph{W}}}_{pq}
\end{pmatrix},
\end{equation}
which we will refer to as the \emph{voltage-current coherency matrix}.
This is analogous to the `visibility matrix', $\mathsf{V}_{pq}$,
of \citet{Smirnov:2011a}.

\section{Two point coherency relationships\label{sec:2pt-coherency}}

Now we have introduced the two-point coherency matrix $\mathbb{B}_{pq}$
and the voltage-current coherency matrix $\mathbb{V}_{pq}$, we can
formulate relationships between the two. In this section, we first
formulate a general coherency relationship describing propagation
from a source of electromagnetic radiation to two spacetime coordinates.
We then show that this relationship underlies both the RIME and the
vC-Z relationship.

\subsection{A general two point coherency relationship}

Suppose we have two sensors, located at points $\mbox{\textbf{\emph{r}}}_{p}$
and $\mbox{\textbf{\emph{r}}}_{q}$, which fully measure all components
of the electromagnetic field. Assuming linearity, the propagation
of an electromagnetic field $\mbox{\textbf{\emph{f}}}_{0}=\begin{pmatrix}\mbox{\textbf{\emph{e}}}(\mbox{\textbf{\emph{r}}}_{0},t) & \mbox{\textbf{\emph{h}}}(\mbox{\textbf{\emph{r}}}_{0},t)\end{pmatrix}^{T}$
from a point $\mbox{\textbf{\emph{r}}}_{0}$ to $\mbox{\textbf{\emph{r}}}_{p}$
and $\mbox{\textbf{\emph{r}}}_{q}$ can be encoded into a 6$\times$6
matrices, $\mathbb{T}_{p}$ and $\mathbb{T}_{q}$:
\begin{align}
\mbox{\textbf{\emph{f}}}_{p} & =\mathbb{T}_{p}\mbox{\textbf{\emph{f}}}_{0}\\
\mbox{\textbf{\emph{f}}}_{q} & =\mathbb{T}_{q}\mbox{\textbf{\emph{f}}}_{0}
\end{align}
The coherency between the two signals $\mbox{\textbf{\emph{f}}}_{p}$
and $\mbox{\textbf{\emph{f}}}_{q}$ is then given by the matrix $\mathbb{B}_{pq}$:

\begin{align}
\mathbb{B}_{pq} & =\left\langle \mbox{\textbf{\emph{f}}}_{p}\otimes\mbox{\textbf{\emph{f}}}_{q}\right\rangle \label{eq:tp_tq_outer}\\
 & =\left\langle (\mathbb{T}_{p}\mbox{\textbf{\emph{f}}}_{0})\otimes(\mathbf{\mathbb{T}}_{q}\mbox{\textbf{\emph{f}}}_{0})\right\rangle \\
 & =\left\langle \mathbf{\mathbb{T}}_{p}(\mbox{\textbf{\emph{f}}}_{0}\otimes\mbox{\textbf{\emph{f}}}_{0})\mathbb{T}_{q}^{H}\right\rangle \\
 & =\mathbf{\mathbb{T}}_{p}\mathbb{B}\mathbf{\mathbb{T}}_{q}^{H}
\end{align}
we can write this in terms of block matrices
\begin{equation}
\begin{pmatrix}\mbox{\textbf{\emph{E}}}_{pq} & \mbox{\textbf{\emph{M}}}_{pq}\\
\mbox{\textbf{\emph{N}}}_{pq} & \mbox{\textbf{\emph{H}}}_{pq}
\end{pmatrix}=\begin{pmatrix}\mbox{\textbf{\emph{A}}}_{p} & \mbox{\textbf{\emph{B}}}_{p}\\
\mbox{\textbf{\emph{C}}}_{p} & \mbox{\textbf{\emph{D}}}_{p}
\end{pmatrix}\begin{pmatrix}\mbox{\textbf{\emph{E}}}_{00} & \mbox{\textbf{\emph{M}}}_{00}\\
\mbox{\textbf{\emph{N}}}_{00} & \mbox{\textbf{\emph{H}}}_{00}
\end{pmatrix}\begin{pmatrix}\mbox{\textbf{\emph{A}}}_{q} & \mbox{\textbf{\emph{B}}}_{q}\\
\mbox{\textbf{\emph{C}}}_{q} & \mbox{\textbf{\emph{D}}}_{q}
\end{pmatrix}^{H}\label{eq:two-pt-coherency}
\end{equation}
This is the most general form that relates the coherency at two points
$\mbox{\textbf{\emph{r}}}_{p}$ and $\mbox{\textbf{\emph{r}}}_{q}$,
to the electromagnetic energy density at point $\mbox{\textbf{\emph{r}}}_{0}$.

In radio astronomy, antennas are used as sensors to measure the electromagnetic
field. Following from Eq.~\ref{eq:two-pt-coherency}, we may write
an equation relating voltage and current coherency: 
\begin{equation}
\mathbb{V}_{pq}=\mathbf{\mathbb{T}}{}_{p}^{'}(\mathbf{\mathbb{T}}_{p}\mathbb{B}\mathbb{T}_{q}^{H})\mathbf{\mathbb{T}}{}_{q}^{'H}.\label{eq:measurement-eq-non-collapsed}
\end{equation}
As the $\mathbb{T}'$ and \textbf{$\mathbb{T}$} matrices are both
$\mbox{6}\times\mbox{6}$ , we can are both collapse these matrices
into one overall matrix. Eq.~\ref{eq:measurement-eq-non-collapsed}
then becomes

\begin{equation}
\mathbb{V}_{pq}=\mathbf{\mathbb{T}}{}_{p}\mathbb{B}\mathbf{\mathbb{T}}_{q}^{H},\label{eq:RIME-basic}
\end{equation}
which is the general form that relates the voltage-coherency matrix
$\mathbb{V}_{pq}$ to the brightness coherency $\mathbb{B}.$ 

Equation\ \ref{eq:RIME-basic} is a central result of this paper.
It is a general case which relates the EM field at a given point in
space-time to the voltage and current coherencies in between pairs
of telescopes. In the sections that follow, we show that generalized
versions of the Van-Cittert-Zernicke theorem and RIME may be formulated based upon this coherency relationship,
and that the common formulations can be derived from these general
results.

\subsection{The Radio Interferometer Measurement Equation}

By comparing Eq.~\ref{eq:jones-rime-2} with Eq.~\ref{eq:two-pt-coherency},
it is apparent that the Jones formulation of the RIME is retrieved
by setting all but the top left block matrices to zero, such that
we have 
\begin{equation}
\mbox{\textbf{\emph{V}}}_{pq}=\mbox{\textbf{\emph{A}}}_{p}\mbox{\textbf{\emph{E}}}_{00}\mbox{\textbf{\emph{A}}}_{q}^{H}.
\end{equation}
But under what assumptions may we ignore the other entries of Eq.~\ref{eq:two-pt-coherency}?
To answer this, we may note that monochromatic plane waves in free
space have \textbf{\emph{E}}\textbf{ }and \textbf{\emph{H}}\textbf{
}are in phase and mutually perpendicular:
\begin{align}
\mbox{\textbf{\emph{e}}}(\mbox{\textbf{\emph{r}}},t) & =\begin{pmatrix}e_{x}(\mbox{\textbf{\emph{r}}},t) & e_{y}(\mbox{\textbf{\emph{r}}},t)\end{pmatrix}^{T}\nonumber\\
\mbox{\textbf{\emph{h}}}(\mbox{\textbf{\emph{r}}},t) & =\frac{1}{c_{0}}\begin{pmatrix}-e_{y}(\mbox{\textbf{\emph{r}}},t) & e_{x}(\mbox{\textbf{\emph{r}}},t)\end{pmatrix}^{T}
\label{eq:MPWZ}
\end{align}
Where $c_{0}$ is the magnitude of the speed of light. In such a case,
all coherency statistics can be derived from the $\mbox{2}\times\mbox{2}$
brightness matrix~$\mathsf{B}$. \citet{Carozzi:2009hf} show that
the field coherencies can be written  
\begin{align}
\mathbb{B} & =\begin{pmatrix}\mbox{\textbf{\emph{E}}}_{pq} & \mbox{\textbf{\emph{M}}}_{pq}\\
\mbox{\textbf{\emph{N}}}_{pq} & \mbox{\textbf{\emph{H}}}_{pq}
\end{pmatrix}=\begin{pmatrix}\mathsf{B} & \mathsf{B}\mbox{\textbf{\emph{F}}}^{T}\\
\mbox{\textbf{\emph{F}}}\mathsf{B} & \mbox{\textbf{\emph{F}}}\mathsf{B}\mbox{\textbf{\emph{F}}}^{T}
\end{pmatrix}\label{eq:Carozzi-matrix}
\end{align}
where $\mathbf{F}$ is the matrix 
\begin{equation}
\mbox{\textbf{\emph{F}}}=\frac{1}{c_{0}}\begin{pmatrix}0 & 1\\
-1 & 0
\end{pmatrix}.
\end{equation}
Under these conditions, the rank of $\mathbb{B}$ is 2, so the relationship
in Eq.~\ref{eq:two-pt-coherency} is over constrained. It follows
that the $\mbox{2}\times\mbox{2}$ RIME is perfectly acceptable ---
and indeed preferable to Eq.~\ref{eq:two-pt-coherency} --- for describing
coherency of plane waves that propagate through free space.

There are numerous situations in which we cannot assume that we have
monochromatic plane waves. This includes near field sources where
the wavefront is not well approximated by a plane wave; propagation
through ionized gas; and situations where we choose not to treat our
field as a superposition of quasi-monochromatic components. Most importantly,
the assumptions that underlie the $\mbox{2}\times\mbox{2}$ RIME do
not hold within the analogue components of a telescope, where the
signal does not enjoy free space impedance.\emph{ }

\subsection{A 2N-port transmission matrix based RIME\label{sub:2N-port-trans-RIME}}

For a dual polarization telescope, a 4-port description (2-in 2-out)
of our analogue system is more appropriate than the general 6-port
description. Using the result \ref{eq:Carozzi-matrix} above and Eq.~\ref{eq:RIME-basic},
we can write a relationship

\begin{equation}
\mbox{\textbf{\emph{V}}}_{pq}=\begin{pmatrix}\mbox{\textbf{\emph{A}}}_{p} & \mbox{\textbf{\emph{B}}}_{p}\\
\mbox{\textbf{\emph{C}}}_{p} & \mbox{\textbf{\emph{D}}}_{p}
\end{pmatrix}\begin{pmatrix}\mathsf{B} & \mathsf{B}\mbox{\textbf{\emph{F}}}^{T}\\
\mbox{\textbf{\emph{F}}}\mathsf{B} & \mbox{\textbf{\emph{F}}}\mathsf{B}\mbox{\textbf{\emph{F}}}^{T}
\end{pmatrix}\begin{pmatrix}\mbox{\textbf{\emph{A}}}_{q} & \mbox{\textbf{\emph{B}}}_{q}\\
\mbox{\textbf{\emph{C}}}_{q} & \mbox{\textbf{\emph{D}}}_{q}
\end{pmatrix}^{H},\label{eq:2n-port-RIME}
\end{equation}
Here, all block matrices have been reduced in dimensions from $\mbox{3}\times\mbox{3}$
to $\mbox{2}\times\mbox{2}$. This version of the RIME retains the
ability to model analogue components, but uses the approximations
of the vC-Z to express $\mathbb{B}$ in terms of the regular $\mbox{2}\times\mbox{2}$
brightness matrix B. The transmission matrices matrices here are similar
to the 2N-port transmission matrices as defined by \citet{BrandaoFaria2002}.%
\footnote{We note that our definition here is the inverse of that of Faria (input
and output are swapped).%
}

The transmission matrices may be broken down into a chain of cascaded components. That is, for
a cascade of $n$ components, we may write the overall transmission matrix as a product of matrices
representing the individual components:

\begin{equation}
\begin{pmatrix}\mbox{\textbf{\emph{A}}}_{p} & \mbox{\textbf{\emph{B}}}_{p}\\
\mbox{\textbf{\emph{C}}}_{p} & \mbox{\textbf{\emph{D}}}_{p}
\end{pmatrix}=\begin{pmatrix}\mbox{\textbf{\emph{A}}}_{np} & \mbox{\textbf{\emph{B}}}_{np}\\
\mbox{\textbf{\emph{C}}}_{np} & \mbox{\textbf{\emph{D}}}_{np}
\end{pmatrix}\cdots\begin{pmatrix}\mbox{\textbf{\emph{A}}}_{1p} & \mbox{\textbf{\emph{B}}}_{1p}\\
\mbox{\textbf{\emph{C}}}_{1p} & \mbox{\textbf{\emph{D}}}_{1p}
\end{pmatrix}
\end{equation}

This is similar to, but more general than, Jones calculus. In the following section we will explore the difference.

\subsection{Limitations of Jones calculus}

Jones calculus essentially asserts two things. Firstly, it asserts that the voltage 2-vector at the output of a 
dual-polarization system $\bmath{v}_p=(v_{p1},v_{p2})^T$ 
is linear with respect to the EMF 2-vector $\bmath{e}=(e_x,e_y)^T$ at the input of the system:
\begin{equation}
\bmath{v}_p = \bmath{J}_p \bmath{e},
\end{equation}
where the Jones matrix $\bmath{J}_p$ describes the voltage transmission properties of the system.
The second assertion is that for a system composed of $n$ components, the effective Jones matrix 
is a product of the Jones matrices of the components:
\begin{equation}
\bmath{v}_p = \bmath{J}_{np} \cdots \bmath{J}_{1p} \bmath{e}.
\end{equation}

With 2N-port transmission matrices, we instead describe the input of the system by the 4-vector $[\bmath{e},\bmath{h}]^T$, 
which for a monochromatic plane wave is equal to 
\begin{equation}
\begin{pmatrix}\bmath{e}\\ \bmath{h}\end{pmatrix} =
(e_x,e_y,-e_y/c_0,e_x/c_0)^T = 
\begin{pmatrix}\bmath{e}\\ \bmath{F}\bmath{e}\end{pmatrix}, 
\end{equation}
and the output of the system is a 4-vector of 2 voltages and 2 currents $(\bmath{v}_p,\bmath{i}_p)^T$, which is
linear with respect to the input:
\begin{equation}
\begin{pmatrix}\bmath{v}_p\\ \bmath{i}_p\end{pmatrix} =
\begin{pmatrix}\bmath{A}_p & \bmath{B}_p\\ \bmath{C}_p & \bmath{D}_p \end{pmatrix} 
\begin{pmatrix}\bmath{e}\\ \bmath{F}\bmath{e}\end{pmatrix}.
\end{equation}
Note that the output voltage is still linear with respect to the input $\bmath{e}$. Indeed, if one is only interested 
in the voltage, the above becomes
\begin{equation}
\bmath{v}_p = (\bmath{A}_p+\bmath{B}_p\bmath{F})\bmath{e},
\end{equation}
i.e. the system has an effective Jones matrix (i.e. a voltage transmission matrix) of
\begin{equation}
\bmath{J}_p = \bmath{A}_p+\bmath{B}_p\bmath{F}.
\end{equation}
However, the Jones formalism breaks down when the system is composed of multiple components. For example, with 2 
components, we may naively attempt to apply Jones calculus, and describe the voltage transmission matrix of the 
system as a product of the components' voltage transmission matrices:
\begin{equation}
\bmath{J}_p = (\bmath{A}_{2p}+\bmath{B}_{2p}\bmath{F})(\bmath{A}_{1p}+\bmath{B}_{1p}\bmath{F}),
\end{equation}
i.e.
\begin{equation}
\bmath{v}_p = (
\bmath{A}_{2p}\bmath{A}_{1p} + \bmath{B}_{2p}\bmath{F}\bmath{A}_{1p} + 
\bmath{A}_{2p}\bmath{B}_{1p}\bmath{F} + \bmath{B}_{2p}\bmath{F}\bmath{B}_{1p}\bmath{F})\bmath{e}.
\label{eq:jones-naive}
\end{equation}
This, however, completely neglects the current transmission properties. Applying the 2N-port transmission matrix 
formalism, we can see the difference:
\begin{equation}
\begin{pmatrix}\bmath{v}_p\\ \bmath{i}_p\end{pmatrix} =
\begin{pmatrix}\bmath{A}_{2p} & \bmath{B}_{2p}\\ \bmath{C}_{2p} & \bmath{D}_{2p} \end{pmatrix} 
\begin{pmatrix}\bmath{A}_{1p} & \bmath{B}_{1p}\\ \bmath{C}_{1p} & \bmath{D}_{1p} \end{pmatrix} 
\begin{pmatrix}\bmath{e}\\ \bmath{F}\bmath{e}\end{pmatrix},
\end{equation}
from which we can derive an expression for the voltage vector:
\begin{equation}
\bmath{v}_p = 
(\bmath{A}_{2p}\bmath{A}_{1p} + \bmath{B}_{2p}\bmath{C}_{1p} + 
\bmath{A}_{2p}\bmath{B}_{1p}\bmath{F} + \bmath{B}_{2p}\bmath{D}_{1p}\bmath{F})\bmath{e},
\label{eq:jones-2N}
\end{equation}
which differs from Eq.~\ref{eq:jones-naive} in the second and fourth term of the sum, since in general
\begin{equation}
\bmath{F}\bmath{A}_{1p} \neq \bmath{C}_{1p},~~\bmath{F}\bmath{B}_{1p} \neq \bmath{D}_{1p}.
\label{eq:jones-difference}
\end{equation}

To summarize, because Jones calculus operates on voltages alone, and ignores impedance matching, 
we cannot use it to accurately represent the voltage response of an analogue system by a product of the voltage 
responses of its individual components. The difference is summarized in Eqs.~\ref{eq:jones-naive}--\ref{eq:jones-difference}; 
a practical example is given in Sect.~\ref{sub:Modelling-real-analogue}. 
By contrast, the 2N-port transmission matrix formalism does allow us to break down the overall system response 
into a product of the component responses, by taking both voltages and currents into account.

Since we have now shown the $2\times2$ form of the RIME to be insufficient, the obvious question arises, why have
we been getting away with using it? Historically, practical applications of the RIME have tended to follow 
the formulation of \citet{JEN:note185}, rolling the electronic response of the overall system (as well as 
tropospheric phase, etc.) into a single `$G$-Jones' term that is solved for during calibration. Under these
circumstances, the $2\times2$ formalism is perfectly adequate -- it is only when we attempt to model the individual 
components of the analogue receiver chain that its limitations are exposed. On the other hand, \citet{Carozzi:2009hf} 
have highlighted the limitations of Jones calculus in the wide-field polarization regime.

\section{Tensor formalisms of the RIME}

Up until now, we have presented our $6\times6$ RIME using matrix
notation. We now briefly discuss how the work presented here is closely
related to the tensor formalism presented in \citet{Smirnov:2011d}. 

As is discussed in \citet{Smirnov:2011d}, the classical Jones formulation
of the RIME is in the vector space $\mathbb{C}^{2}$. The formulation
proposed by \citet{Carozzi:2009hf} is instead in $\mathbb{C}^{3}$.
In contrast, our Eq.\ \ref{eq:RIME-basic} can be considered to work
in $\mathbb{C}^{6}$; that is, our EMF vector has 6 components: 
\begin{eqnarray}
e^{i} & = & \sum_{j=1}^{3}e_{j}x^{j}+\sum_{k=1}^{3}h_{j}x^{j}\equiv\sum_{j=1}^{6}e_{j}x^{j}.
\end{eqnarray}
The coherency of two voltages is once again defined via the outer
product $e^{i}\bar{e}_{j}$, quite remarkably giving a (1,1)-type
tensor expression identical Eq.\ 9 of \citet{Smirnov:2011d}:
\begin{equation}
\mbox{\emph{V}}_{qj}^{pi}=\mbox{\emph{J}}_{\alpha}^{pi}\mbox{\emph{B}}_{\beta}^{\alpha}\bar{\mbox{\emph{J}}}_{qj}^{\beta}.
\end{equation}
In the next section, we show an alternative RIME based upon a (2,0)-type tensor commonly encountered in special relativity. 

\subsection{A relativistic RIME}

Another potential reformulation of the RIME involves the electromagnetic tensor of special relativity. The classical RIME formulation
implicitly assumes that both antennas measure the EMF in the same inertial reference frame. If this is not the 
case (consider, e.g., space VLBI), then we must in principle account for the fact that the observed EMF is 
altered when moving from one reference frame to another, and in particular, that the $\bmath{e}$ and $\bmath{h}$ 
components intermix. In special relativity, this can be elegantly formulated in terms of the \emph{electromagnetic 
field tensor}, which represents the 6 independent components of the EMF by a (2-0)-type tensor:

\begin{equation}
F^{\alpha\beta}=\left(\begin{array}{cccc}
0 & -e_{x}/c & -e_{y}/c & -e_{z}/c\\
e_{x}/c & 0 & -h_{z} & h_{y}\\
e_{y}/c & h_{z} & 0 & -h_{x}\\
e_{z}/c & -h_{y} & h_{x} & 0
\end{array}\right),
\label{eq:F-general}
\end{equation}

The advantage of this formulation is that the EMF tensor follows standard coordinate transform laws of special 
relativity. That is, for a different inertial reference frame given by the Lorentz transformation tensor 
$\Lambda^\alpha_{\alpha'}$, the EMF tensor transforms as:
\begin{equation}
F'^{\alpha\beta} = \Lambda^\alpha_{\alpha'} \Lambda^\beta_{\beta'} F^{\alpha'\beta'}. 
\end{equation}
The measured 2-point coherency between $[F_p]$ and $[F_q]$ can be formally defined as the average of the 
outer product:
\begin{equation}
[V_{pq}]^{\alpha\beta\gamma\delta} = 2 c^2 \langle [F_p]^{\alpha\beta}[F^H_q]^{\gamma\delta} \rangle, 
\end{equation}
where $\cdot^H$ represents the conjugate tensor, i.e. $[F^H]^{\gamma\delta}=\bar{F}^{\delta\gamma}$. A factor of 2
is introduced for the same reasons as in \citet{Smirnov:2011a}, and the reason for $c^2$ will be apparent below. 
Note that the indices in the brackets should be treated as labels, while those outside the brackets are proper tensor indices. 

Let us now pick a reference frame for the signal (`frame zero'), and designate the EMF tensor in that frame 
by $[F_0]$, or $[F_0(\bmath{\bar{x}})]$ to emphasize that this is a function of the four-position 
$\bmath{\bar{x}}=(ct,\bmath{x})$. The $[F_0(\bmath{\bar{x}})]$ field follows Maxwell's equations; in the case
of a monochromatic plane wave propagating along direction $\bmath{\bar{z}}=(1,\bmath{z})$, this has a particularly simple solution
of 
\begin{equation}
[F_0(\bmath{\bar{x}})] = [F_0(\bmath{\bar{x}}_0)]e^{-2\pi i \lambda^{-1} (\bmath{\bar{x}}-\bmath{\bar{x}}_0)\cdot \bmath{\bar{z}}}.
\label{eq:F0-x}
\end{equation}

Let us now consider two antennas located at $\bmath{p}$ and $\bmath{q}$. The 2-point coherency measured in frame zero 
becomes
\begin{eqnarray}
[V_{pq}]^{\alpha\beta\gamma\delta} & = & 
  2 c^2 \langle [F_0(\bmath{\bar{p}})]^{\alpha\beta}[F^H(\bmath{\bar{q}})]^{\gamma\delta} \rangle \nonumber\\
&=&  K_p \left [ 2 c^2 \langle [F_0(\bmath{\bar{x}}_0)]^{\alpha\beta}[F^H(\bmath{\bar{x}}_0)]^{\gamma\delta} \rangle \right ] K^H_q,
\end{eqnarray}
where $K_p$ is the complex exponent of Eq.~\ref{eq:F0-x}, and is the direct equivalent of the $K$-Jones 
term of the RIME \citep{Smirnov:2011a}. The quantity in the square brackets is the equivalent of the
brightness matrix $\mathsf{B}$, which we'll call the \emph{brightness tensor}:

\begin{equation}
B^{\alpha\beta\gamma\delta} = [B_0]^{\alpha\beta\gamma\delta} = 2c^2 \langle [F_0(\bmath{\bar{x}}_0)]^{\alpha\beta}[F_0^H(\bmath{\bar{x}}_0)]^{\gamma\delta} \rangle.
\end{equation}
Each element of the brightness tensor gives the coherency between two components of the EMF observed in the chosen
reference frame (`frame zero'). Nominally, the brightness tensor has $4^4=256$ components, but only 36 are unique 
and non-zero (given the 6 components of the EMF). \citet{Carozzi:2006bj} show that the brightness tensor may be 
decomposed into a set of antisymmetric second rank tensors (`sesquilinear-quadratic tensor concomitants') 
that are irreducible under Lorentz transformations. The physical interpretation of the 36 unique quantities within 
the brightness matrix is discussed in \citet{Bergman:2008p7859}, with regards to the aforementioned irreducible tensorial set.

While the brightness tensor has redundancy not present in the tensorial set of \citet{Carozzi:2006bj}, we shall continue to use it
as a basis to our relativistic RIME for clarity of analogy to the brightness coherency matrix of Eq.~\ref{eq:RIME-basic}, and as
it leads to a relativistic RIME for which we can define transformation tensors analogous to Jones matrices. The redundancy 
can be described by a number of symmetry properties of the brightness tensor: it is 
(a) \emph{Hermitian} with respect to swapping the  first and second pair of indices:
\begin{equation}
B^{\alpha\beta\gamma\delta} = \bar{B}^{\gamma\delta\alpha\beta},
\label{eq:BT-herm}
\end{equation}
(b) \emph{antisymmetric} within each index pair (since the EMF tensor itself is antisymmetric, i.e. $F^{\alpha\beta}=-F^{\beta\alpha}$):
\begin{equation}
B^{\alpha\beta\gamma\delta} =  -B^{\beta\alpha\gamma\delta} = -B^{\alpha\beta\delta\gamma}
\label{eq:BT-antisym}
\end{equation}
To see the direct analogy to the brightness matrix, consider again the case of the monochromatic plane wave propagating 
along $\bmath{z}$ (Eq.~\ref{eq:MPWZ}). The EMF tensor then takes a particularly simple form:
\begin{equation}
\label{eq:F-planewave}
F^{\alpha\beta} = \frac{1}{c} \left(\begin{array}{cccc}
0       & -e_{x} & -e_{y} & 0 \\
e_{x} & 0        & 0        & e_x \\
e_{y} & 0        & 0        & e_y \\
0       & -e_{x} & -e_y   & 0
\end{array}\right),
\end{equation}
and the brightness tensor has only $8^2=64$ non-zero components, with the additional `0-3' symmetry property:
\begin{eqnarray}
B^{0\beta\gamma\delta} = B^{3\beta\gamma\delta} && B^{\alpha0\gamma\delta} = B^{\alpha3\gamma\delta} \nonumber\\
B^{\alpha\beta0\delta} = B^{\alpha\beta3\delta} && B^{\alpha\beta\gamma0} = B^{\alpha\beta\gamma3} 
\label{eq:BT-sym}
\end{eqnarray}
Four unique components can be defined in terms of the Stokes parameters:
\begin{eqnarray}
B^{0110} = I+Q && B^{0220} = I-Q \nonumber\\ 
B^{0120} = U+iV && B^{0210} = U-iV 
\label{eq:BT-IQUV}
\end{eqnarray}
and conversely,
\begin{eqnarray}
I = \frac{B^{0110}+B^{0220}}{2} && Q = \frac{B^{0110}-B^{0220}}{2}, \nonumber\\
U = \frac{B^{0120}+B^{0210}}{2} && V = \frac{B^{0120}-B^{0210}}{2i}. 
\label{eq:IQUV-BT}
\end{eqnarray}
The other non-zero components of the brightness tensor can be derived using the Hermitian, antisymmetry and 
0-3 symmetry properties. Finally, for an unpolarized plane wave, only 32 components of the brightness tensor
are non-zero and equal to $\pm I$. As these will be useful in further calculations, they are summarized in Table~\ref{tab:BT}.

\begin{table}
\begin{centering}
\begin{tabular}{c|rrrrrrrr}
   & 01 & 02 & 10 & 13 & 20 & 23 & 31 & 32 \\
\hline
01 & $-I$ &    & $I$ & $I$ &    &    & $-I$ &    \\  
02 &    & $-I$ &    &    & $I$ & $I$ &    & $-I$ \\  
10 & $I$ &    & $-I$ & $-I$ &    &    & $I$ &    \\  
13 & $I$ &    & $-I$ & $-I$ &    &    & $I$ &    \\  
20 &    & $I$ &    &    & $-I$ & $-I$ &    & $I$ \\  
23 &    & $I$ &    &    & $-I$ & $-I$ &    & $I$ \\  
31 & $-I$ &    & $I$ & $I$ &    &    & $-I$ &    \\  
32 &    & $-I$ &    &    & $I$ & $I$ &    & $-I$ \\  
\end{tabular}
\end{centering}
\caption{The non-zero components of the brightness tensor $B^{\alpha\beta\gamma\delta}$ for an unpolarized
plane wave. Rows correspond to $\alpha\beta$, columns to $\gamma\delta$.}
\label{tab:BT}
\end{table}

So far this has been nothing more than a recasting of the RIME using EMF tensors. Consider, however, the case where antennas $p$ and $q$ measure the signal in different inertial reference frames. The EMF tensor observed by
antenna $p$ becomes
\begin{equation}
[F_p]^{\alpha\beta} = [\Lambda_p]^\alpha_{\alpha'} [\Lambda_p]^\beta_{\beta'} [F_0]^{\alpha'\beta'},
\end{equation}
where $[\Lambda_p]^\alpha_\mu$ is the Lorentz tensor corresponding to the transform between the signal frame and the
antenna frame. Since the same Lorentz tensor always appears twice in these equations (due to $F$ being a (2,0)-type tensor), let us designate
\begin{equation}
[\Lambda_p]^{\alpha\beta}_{\alpha'\beta'} = [\Lambda_p]^\alpha_{\alpha'} [\Lambda_p]^\beta_{\beta'}
\label{eq:LL}
\end{equation}
for compactness. The measured coherency now becomes
\begin{equation}
[V_{pq}]^{\alpha\beta\gamma\delta} = 
K_p {[\Lambda_p]}^{\alpha\beta}_{\alpha'\beta'} 
[B_0]^{\alpha'\beta'\gamma'\delta'}
[\Lambda_q]^{\gamma\delta}_{\gamma'\delta'} 
K^H_q.
\label{eq:RRIME-KL}
\end{equation}
The equivalent of
Jones matrices would be (2,2)-type `Jones tensors' $J^{\alpha\beta}_{\alpha'\beta'}$, so a more general formulation of the 
above would be
\begin{equation}
[V_{pq}]^{\alpha\beta\gamma\delta} = 
{[J_p]}^{\alpha\beta}_{\alpha'\beta'} 
[B_0]^{\alpha'\beta'\gamma'\delta'}
[J^H_q]^{\gamma\delta}_{\gamma'\delta'},
\label{eq:RRIME-J}
\end{equation}
where tensor conjugation $\cdot^H$ is defined as:
\begin{equation}
[J^H]^{\alpha\beta}_{\alpha'\beta'}=\bar{J}^{\beta\alpha}_{\beta'\alpha'}.
\end{equation}
Note that both the $K$ and $\Lambda$ terms of Eq.~\ref{eq:RRIME-KL} can be considered as special 
examples of Jones tensors. The $K$ term can be explicitly written as a (2,2)-type tensor via two Kronecker deltas:
\begin{equation}
[K_p]^{\alpha\beta}_{\alpha'\beta'} = K_p \delta^{\alpha}_{\alpha'}\delta^{\beta}_{\beta'},
\end{equation} 
while the $\Lambda$ term is a (2,2)-type tensor by definition (Eq.~\ref{eq:LL}), noting that
\begin{equation}
[\Lambda^H_p]^{\alpha\beta}_{\alpha'\beta'} = [\Lambda_p]^{\alpha\beta}_{\alpha'\beta'},
\end{equation}
since the components of any Lorentz transformation tensor $\Lambda^{\alpha}_{\alpha'}$ are real. 

Finally, let us note the equivalent of the `chain rule' for Jones tensors. If the signal chain is represented by 
a sequence of Jones tensors $[J_{p,n}],...,[J_{p,1}]$ (including $K$ terms, $\Lambda$ terms, and all
instrumental and propagation effects), then the overall response is given by
\begin{equation}
[J_p]^{\alpha\beta}_{\alpha'\beta'} = 
[J_{p,n}]^{\alpha\beta}_{\alpha_n\beta_n}
[J_{p,n-1}]^{\alpha^n\beta^n}_{\alpha_{n-1}\beta_{n-1}}
\cdots
[J_{p,1}]^{\alpha^2\beta^2}_{\alpha'\beta'}.
\label{eq:JT-chain}
\end{equation}
Equations~\ref{eq:RRIME-KL}, \ref{eq:RRIME-J} and \ref{eq:JT-chain} above constitute a relativistic reformulation of the 
RIME (RRIME).  The RRIME allows us to incorporate relativistic effects into our measurement equation. 
That is, we can treat relativistic motion as an `instrumental' effect. Some interesting observational consequences, and
the relation of the RRIME to work from other fields are treated in the discussion below.

\section{Discussion}

The formulations presented here highlight the relationship between the seemingly disparate 
fields of microwave networking and special relativity to measurements in radio astronomy. This is a 
remarkable illustration of how these fields are intrinsically related by underlying fundamental physics. 

Indeed, it has 
long been known that Jones and Mueller transformation matrices are  
related to Lorentz transformations by the Lorentz group. 
Wiener was aware as early as 1928 that the the 2$\times$2 coherency matrix could be written in terms of the 
Pauli spin matrices \citep{wiener1928, wiener1930}. More recently, \citet{Baylis:1993} presents a more general geometric algebra formalism 
that unifies Jones and Mueller matrices with Stokes parameters and the Poincar\'{e} sphere. 
\citet{Han:1997jones} show that the Jones formalism is a representation of the six-parameter 
Lorentz group; further to this \citet{Han:1997stokes} show that the Stokes parameters form a Minkowskian 
four-vector, similar to the energy-momentum four-vector in special relativity. 

In contrast, the RRIME presented here is novel as it fully describes interferometric measurement between
two points, instead of simply describing polarization states and defining a transformation algebra. Similarly, the 2N-port
RIME specifically shows how microwave networking methods can be incorporated into a ME.

What do we gain from using these more general measurement equations in lieu of the simpler 2$\times$2 RIME?
For most applications, it will suffice to simply be aware of the standard RIME's limitations, 
and to work in a piecemeal fashion. For example, one can quite happily use Jones matrices to describe 
free space propagation, and microwave network methods  to describe discrete analogue components. 
An overall `system Jones' matrix may be derived to describe instrumental effects, but this matrix should never
be decomposed into a Jones chain. 

Similarly, special relativity describes relativistic effects through Lorentz transformations 
acting upon the EMF tensor.  We can treat relativistic motion as an instrumental effect by using the RRIME,
or we can apply special relativistic corrections separately, as required. The effect of relativistic boosts on 
the Stokes parameters are considered in \citet{cocke:1972}. 

In the subsections that follow, we present some potential use cases that highlight the usefulness of the 2N-port 
and relativistic RIME formulations.

\subsection{Modelling real analogue components\label{sub:Modelling-real-analogue}}

The 2N-port RIME may have practicality in absolute flux calibration of radio telescopes.
Generally, interferometer data is calibrated by assuming that the sky brightness is known,
or at least approximately known. If we wish to do absolute calibration of a telescope without
making assumptions about the sky, we must instead make sure that we accurately model the
analogue components of our telescope. A particularly ubiquitous method of device
characterization within microwave engineering is the use of scattering parameters; we briefly
introduce these below before incorporating them into the 2N-port RIME.

\subsubsection{Scattering parameters}

Voltage, current and impedance are somewhat abstract concepts at microwave
frequencies, so engineers often use scattering parameters to quantify
a device's characteristics. Scattering parameters relate the incident
and reflected voltage waves on the ports of a microwave network. The
scattering matrix, \textbf{\emph{S}}, is given by 
\begin{equation}
\begin{pmatrix}v_{1}^{-}\\
v_{2}^{-}\\
\vdots\\
v_{n}^{-}
\end{pmatrix}=\begin{pmatrix}S_{11} & S_{12} & \cdots & S_{1n}\\
S_{21} &  &  & \vdots\\
\vdots &  & \ddots & \vdots\\
S_{n1} & \cdots & \cdots & S_{nn}
\end{pmatrix}\begin{pmatrix}v_{1}^{+}\\
v_{2}^{+}\\
\vdots\\
v_{n}^{+}
\end{pmatrix},
\end{equation}
where $v_{n}^{+}$ is the amplitude of the voltage wave incident on
port $n$, and $v_{n}^{-}$ is the amplitude of the voltage wave reflected
from port $n$. 

For a dual polarization system (we will label the polarization \emph{x}
and \emph{y}) with negligible crosstalk, we can model the analogue
chain for each polarization as a discrete 2-port network. Assuming
that the analogue chains have the same number of components (but not
that the components are identical), the transmission matrix for each
pair of components is 
\begin{equation}
\mbox{\textbf{\emph{T}}}=\begin{pmatrix}\tilde{A}_{x} & 0 & \tilde{B}_{x} & 0\\
0 & \tilde{A}_{y} & 0 & \tilde{B}_{y}\\
\tilde{C}_{x} & 0 & \tilde{D}_{x} & 0\\
0 & \tilde{C}_{y} & 0 & \tilde{D}_{y}
\end{pmatrix},
\end{equation}
where the elements are from the \emph{ABCD }matrices of the \emph{x
}and \emph{y }polarizations, and are given by the relations 
\begin{align}
\tilde{A} & =\frac{1+S_{12}S_{21}+S_{22}-S_{11}(1+S_{22})}{2S_{12}}\label{eq:s-to-t-first}\\
\tilde{B} & =-Z_{0}\frac{1+S_{11}-S_{12}S_{21}+S_{22}+S_{11}S_{22}}{2S_{12}}\\
\tilde{C} & =\frac{1}{Z_{0}}\frac{-1+S_{11}+S_{12}S_{21}+S_{22}-S_{11}S_{22}}{2S_{12}}\\
\tilde{D} & =\frac{1+S_{11}+S_{12}S_{21}-S_{22}-S_{11}S_{22}}{2S_{12}}\label{eq:s-to-t-last}
\end{align}
Here, $Z_{0}$ is the characteristic impedance of the analogue chain,
which in most telescopes is set to 50 or 75 ohms. We have added tildes
to the \emph{ABCD }parameters, as we are using the inverse definition
to that in \citet{Pozar2005}. 

If the system has significant crosstalk between polarizations, the
analogue chain is more accurately modelled as a 4-port network. In
this case the relationships are not so simple, and the off-diagonal
entries of the block matrices of \textbf{\emph{T}}\textbf{ }will no
longer be zero.

\subsubsection{Scattering matrix example}

We now present a simple illustration of the differences between assigning
a component a scalar value (as is done in the Jones formalism), and
by modelling it as a 2-port network. Consider a component with a scattering
matrix

\begin{equation}
\mbox{\textbf{\emph{S}}}=\begin{pmatrix}0.1\angle0^{\circ} & 0.9\angle0^{\circ}\\
0.9\angle0^{\circ} & 0.1\angle0^{\circ}
\end{pmatrix}
\end{equation}
for a given quasi-monochromatic frequency. Here, values are presented in angle notation
to emphasize that they are complex valued. In decibels, the $S_{11}$
and $S_{22}$ parameters have a magnitude of -10~dB, and the $S_{12}$
and $S_{21}$ parameters have a magnitude of about $-0.5$~dB. Now
suppose we have three identical copies of this component, and we connect
them together in cascade. If one only considered the forward gain
($S_{21}$), one would arrive at an overall $S_{\rm{21tot}}$ of
\begin{equation}
S_{\rm{21tot}}=S_{21}S_{21}S_{21}=0.729
\end{equation}
In contrast, using the standard microwave engineering methods, we
can form a transmission matrix (using equations \ref{eq:s-to-t-first}-\ref{eq:s-to-t-last}
above), and then convert this back into an overall matrix, $\mbox{\textbf{\emph{S}}}_{\rm{tot}}$.
By doing this, one finds
\begin{equation}
\mbox{\textbf{\emph{S}}}_{\rm{cas}}=\begin{pmatrix}0.25\angle0^{\circ} & 0.75\angle0^{\circ}\\
0.75\angle0^{\circ} & 0.25\angle0^{\circ}
\end{pmatrix}.
\end{equation}
The difference becomes more marked as $S_{11}$ and $S_{22}$ increase;
as $S_{11}$ and $S_{22}$ approach zero, the two methods converge. 

When designing a component, $S_{11}$ and $S_{22}$ are generally
optimised to be as small as possible over the operational bandwidth%
\footnote{A notable exception is RF filters, for which $S_{11}$ is often close
to unity out-of-band. In such cases $S_{11}$ is strongly dependent
upon frequency. 
}. Nevertheless, their affect on the overall system is often non-negligible.

\subsubsection{Absolute calibration experiments}

\begin{figure}
\begin{centering}
\includegraphics[width=0.99\columnwidth]{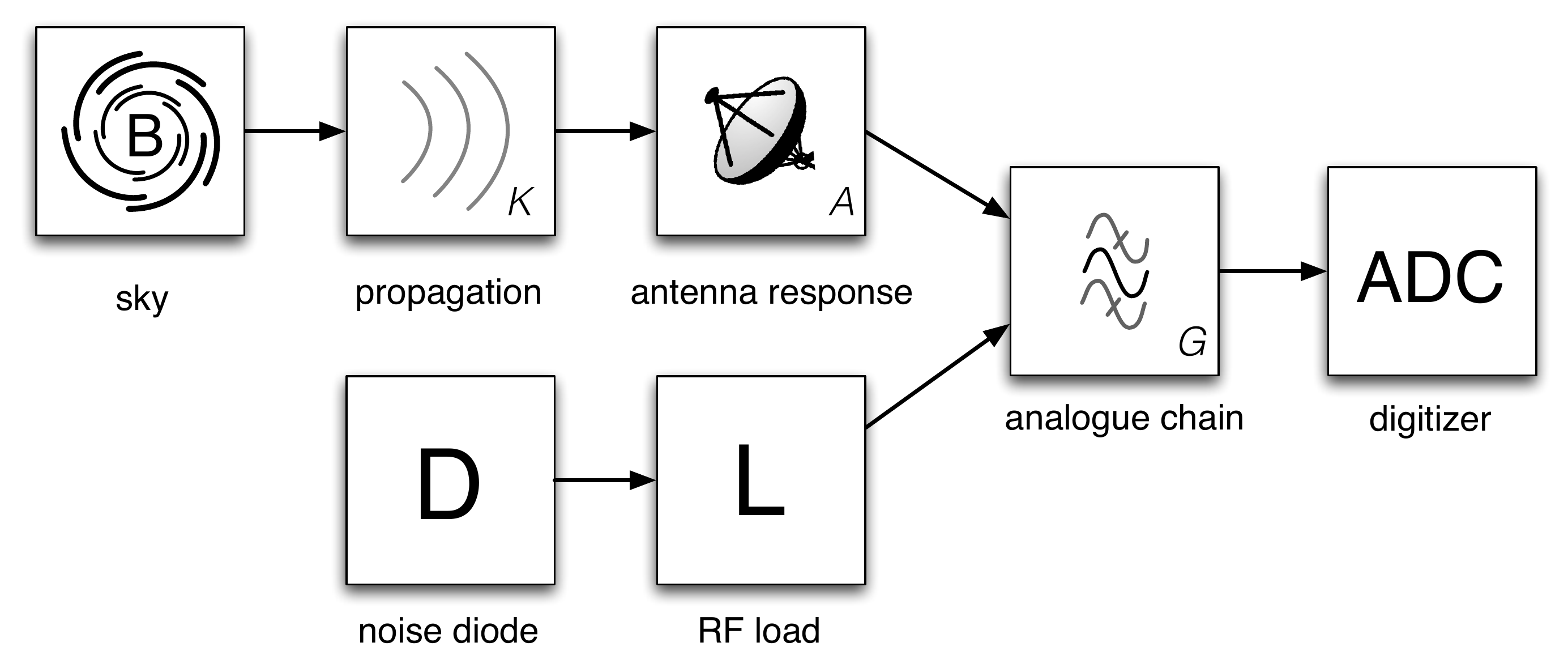}
\par\end{centering}

\protect\caption{Block diagram of a three-state switching experiment. In addition to the antenna path, 
an RF load can in series with a noise diode can be selected. The noise diode can be turned on and off, giving
three possible states: antenna, load, and diode + load. These three states are used for instrumental calibration.
 \label{fig:three-switch}}
\end{figure}

Now we have shown how scattering parameters can be used within the 2N-port RIME, we turn our attention to 
the challenges of absolute calibration of radio telescopes. Measurement of the absolute flux of radio sources is fiendishly hard; 
as such, almost all calibrated flux data presented in radio astronomy literature are pinned to the flux scale presented 
in \citet{Baars:1977}, either directly or indirectly \citep{Kellermann:2009}. Recent experiments, such as the Experiment to Detect the Global EoR Step (EDGES, \citealt{Bowman:2010}),
and the Large-Aperture Experiment to detect the Dark Ages (LEDA, \citealt{leda2012}), are seeking to make precision measurement of the sky temperature (i.e. total power of the
sky brightness) as a function of frequency. The motivation for this is to detect predicted faint (mK) spectral features imprinted
on the sky temperature due to coupling between the microwave background and neutral Hydrogen in the early universe.
For such instruments --- and other instruments with wide field-of-views --- this flux `bootstrapping' method is not sufficient.
 
For experiments such as EDGES a thorough understanding of the analogue systems is
vital to control the systematics that confound calibration. The calibration strategy used for EDGES is detailed in 
\citet{Rogers:2012hd}, and consists of a novel three-state switching and rigorous scattering parameter measurements
with a Vector Network Analyzer (VNA) to remove the bandpass and characterize the antenna.

Our 2N-port RIME allow for scattering parameters to be directly incorporated into a measurement equation that relates
sky brightness to measured voltages at the digitizer. This could be used either to (a) infer the scattering parameters of 
a device given a known sky brightness, or (b) infer the sky brightness from precisely measured scattering parameters. 
The EDGES approach is the latter, but via an ad-hoc method without formal use of a ME. 

\subsubsection{A three-state switching measurement equation}

\begin{figure*}
\subfloat[ \label{fig:leda-shc}.]{\includegraphics[width=0.99\columnwidth]{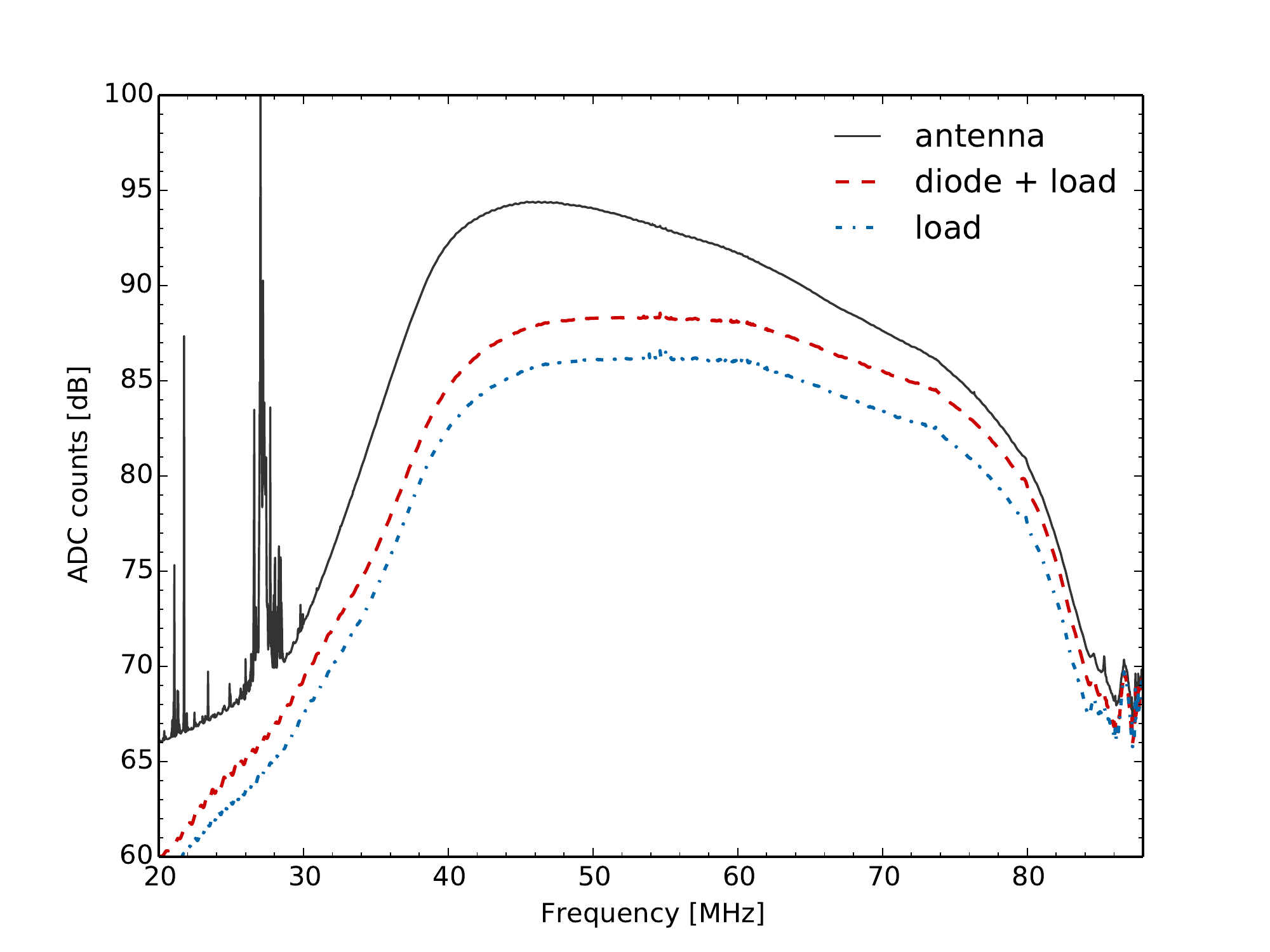}

}\subfloat[  ]{\includegraphics[width=0.99\columnwidth]{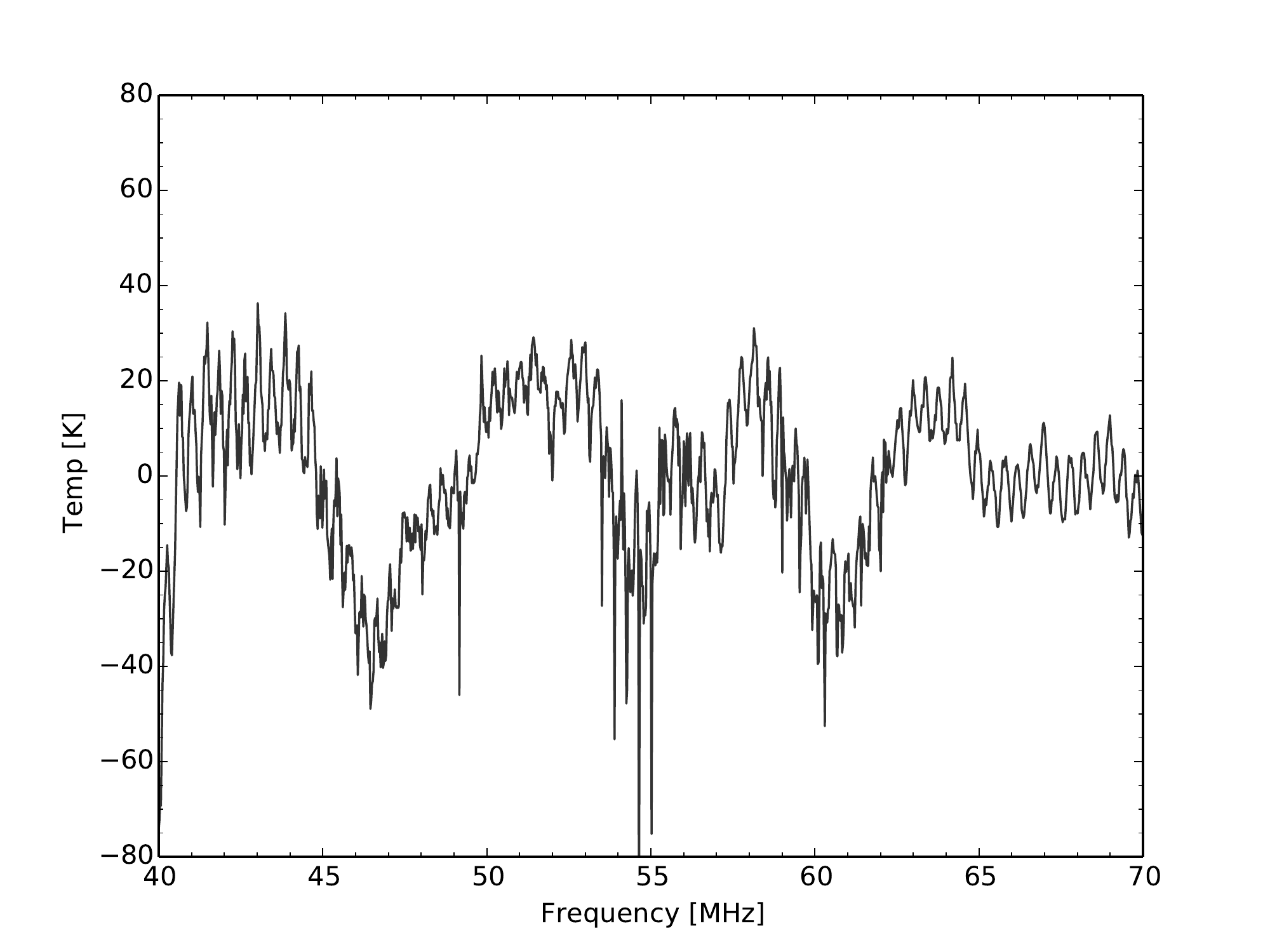}}

\protect\caption{ (a) Data from a LEDA three-state switched antenna during a testing campaign in December, 2013. These correspond to 
the $P_A$, $P_L$, and $P_D$ of Eqns.~\ref{eq:pow-ant}-\ref{eq:pow-diode}.
(b) Residuals after applying Eq.~\ref{eq:ant-temp}, to the data shown in Fig.~\ref{fig:leda-shc} and removing an antenna model power-law fit.
The fast varying sinusoid can be attributed standing waves along the $\sim$300 m of coaxial cable connecting the antenna to the back-end electronics; the sharp negative spikes are due to radio interference.\label{fig:leda-all}}

\end{figure*}

The three-state switching as described in \citet{Rogers:2012hd} involves switching between the antenna  
and a reference load and noise diode, and measuring the resultant power spectra 
($P_{A}$, $P_{L}$, and $P_{D}$, respectively). This is shown as a block diagram in Figure~\ref{fig:three-switch}.
The power as measured in each state is then given by:
\begin{eqnarray}
P_{A} & = & Gk_{B}\Delta\nu(T_{A}+T_{\rm{rx}})\label{eq:pow-ant}\\
P_{L} & = & Gk_{B}\Delta\nu(T_{L}+T_{\rm{rx}})\\
P_{D} & = & Gk_{B}\Delta\nu(T_{D}+T_{L}+T_{\rm{rx}}\label{eq:pow-diode})
\end{eqnarray}
where $k_B$ is the Boltzmann constant, $T_D$ and $T_L$ are the diode and reference load noise temperatures,
$T_{\rm{rx}}$ is the receiver's noise temperature, $G$ is the system gain, 
and $\Delta\nu$ is bandwidth. One can recover the antenna temperature $T_{A}$ by 
\begin{equation}
T_{A}=T_{D}\frac{P_{A}-P_{L}}{P_{D}-P_{L}}+T_{L}\label{eq:ant-temp},
\end{equation}
where the diode and load temperatures, $T_D$ and $T_L$, are known. Antenna temperature is then related to the
sky temperature by
\begin{equation}
T_{\rm{sky}}=T_{A}(1-|\Gamma|^2),\label{eq:sky-ant}
\end{equation}
where the reflection coefficient $\Gamma\equiv S_{11}$. Failure to account for reflections (i.e. impedance mismatch)
results in an unsatisfactory calibration, due to standing waves within the coaxial cable that manifest as 
a sinusoidal ripple on $T_A = T_{A}(\nu)$. This effect is a prime example why one must not use Jones matrices
to describe analogue components separately; an example showing a standing wave present on three-state switch
calibrated spectrum from a prototype LEDA antenna is shown in Figure~\ref{fig:leda-all}.

We may instead write the equations above in terms of 2N-port transmission matrices, and form MEs for the 
three states:
\begin{eqnarray}
\mathbf{V}_{A} & = & \mathbf{G}(\mathbf{T}_A\mathbb{B}_{\rm{sky}}\mathbf{T}_A^{H} + \mathbb{B}_{\rm{rx}})\mathbf{G}^{H}\label{eq:switchmat1}\\
\mathbf{V}_{L} & = & \mathbf{G}(\mathbb{B}_{L} + \mathbb{B}_{\rm{rx}})\mathbf{G}^{H}\\
\mathbf{V}_{D} & = & \mathbf{G}(\mathbf{T}_L\mathbb{B}_{D}\mathbf{T}_L^{H}+\mathbb{B}_{L} + \mathbb{B}_{\rm{rx}})\mathbf{G}^{H}\label{eq:switchmat3}.
\end{eqnarray}
Here, we have replaced the scalar powers $P_{A,L,D}$ with corresponding voltage-coherency matrices $\mathbf{V}_{A,L,D}$, 
temperatures $T_{A,L,\rm{sky},\rm{rx}}$ are replaced with brightness matrices $\mathbb{B}_{A,L,\rm{sky},\rm{rx}}$, and $Gk_B$ is instead represented by 
a system gain matrix $\mathbf{G}$. We have added a transmission matrix for the antenna $\mathbf{T}_A$, and a transmission
matrix $\mathbf{T}_L$ for the load in series with the noise diode. Note that the relation of Eq.~\ref{eq:sky-ant} is now encoded into the 
ME by the matrix $\mathbf{T}_A$, and that we have dropped antenna number subscripts as \emph{p}=\emph{q} 
for autocorrelation measurements.  

It is immediately apparent that the cancellations that occur in the ratio of Eq.~\ref{eq:ant-temp} will not in general occur for 
the equivalent ratio $\mathbf{R} = (\mathbf{V}_{A}-\mathbf{V}_{L})(\mathbf{V}_{D}-\mathbf{V}_{L})^{-1}$. We can however
retrieve the result of Eq.~\ref{eq:ant-temp} by treating the two polarizations separately, setting $\mathbf{G} = G\mathbf{I}$
and $\mathbf{T}_L = \mathbf{I}$, with $\mathbf{T}_A = T_{A}(1-|\Gamma|^2)\mathbf{I}$. 

Our equations Eq.~\ref{eq:switchmat1}-\ref{eq:switchmat3} allow for both polarizations to be treated together, which will be 
important if cross-polarization terms are non-negligible. Also, we may expand Eq.~\ref{eq:switchmat1} to include ionospheric
effects. This ability to combine all effects into a single ME may simplify data analysis and improve calibration accuracy for such experiments.

\subsection{RRIME and Lorentz boosts}

To illustrate how the RRIME can be used to describe relativistic effects, let us first consider the `simple' case of a
relativistically moving source. Suppose we have an unpolarized point source (Table~\ref{tab:BT}), with antennas $p$ and 
$q$ located in the $xy$ plane perpendicular to the direction of propagation $z$ (so the $K$ component becomes unity), both
moving parallel to the $x$ axis with velocity $v$. The Lorentz transformation tensor from the signal frame to the antenna
frame is then

\begin{equation}
[\Lambda_p]^\alpha_{\alpha'}=
\left(\begin{array}{cccc}
\gamma & -\beta\gamma & 0 & 0 \\
-\beta\gamma & \gamma & 0 & 0 \\
0 & 0 & 1 & 0 \\
0 & 0 & 0 & 1 
\end{array}\right),
\label{eq:Lboost}
\end{equation}
where
\begin{equation}
\beta = \frac{v}{c},~~\gamma=\frac{1}{\sqrt{1-\beta^2}}.
\end{equation}
The Lorentz factor $\gamma$ is unity 1 at $v=0$, and goes to infinity with $v\rightarrow c$. This case can be analyzed without
invoking coherencies. Consider the EMF, which in the signal frame is a plane wave (Eq.~\ref{eq:F-planewave}). In the antenna frame it becomes

\begin{equation}
[F_p]^{\alpha\beta} = \frac{1}{c} 
\left(\begin{array}{cccc}
0 				& -e_x 					& -\gamma e_y 	  	& -\beta \gamma e_x \\
e_x 			& 0 					& \beta\gamma e_y 	& \gamma e_x \\
\gamma e_y 		& -\beta\gamma e_y 		& 0 				& e_y \\
\beta\gamma e_x & -\gamma e_x 			& -e_y 				& 0
\end{array}\right ) .
\end{equation} 
Noting that 
\begin{equation}
\left(\begin{array}{ccc}
\gamma^{-1} & 0 & \beta \\
0 & 1 & 0 \\
-\beta & 0 & \gamma^{-1}
\end{array}\right )
\left(\begin{array}{c}
e_{x} \\
\gamma e_y \\
\beta \gamma e_x 
\end{array}\right ) = 
\gamma \left(\begin{array}{c}
e_x \\
e_y \\
0
\end{array}\right ),
\end{equation}
and
\begin{equation}
\left(\begin{array}{ccc}
\gamma^{-1} & 0 & \beta \\
0 & 1 & 0 \\
-\beta & 0 & \gamma^{-1}
\end{array}\right )
\left(\begin{array}{c}
-e_y \\
\gamma e_x \\
-\beta \gamma e_y 
\end{array}\right ) = 
\gamma \left(\begin{array}{c}
-e_y \\
e_x\\
0
\end{array}\right ),
\end{equation}
and $\gamma^{-2}+\beta^2=1,$ we can see that the EMF in the antenna frame is equivalent to a boost of the original EMF 
by $\gamma$ (also called {\em Doppler boost}), coupled to a rotation through $\phi=-\cos^{-1}\beta$ in the $xz$ plane (i.e. the
direction of propagation appears to change, also called {\em relativistic aberration}). Both effects
are well-understood in the guise of {\em relativistic beaming}, and explain why, for example, some AGNs exhibit asymmetric 
jets \citep[see e.g.][]{Sparks1992}.

From an RRIME point of view, a more interesting case arises when one antenna is moving with respect to the other (as is the
case in space VLBI). Let's consider antenna $p$ to be at rest with respect to the 
signal frame (so the $[\Lambda_p]^{\alpha\beta}_{\alpha'\beta'}$ component becomes the equivalent of unity -- the 
product of two Kronecker deltas -- and thus may be dropped), and antenna to be $q$ moving parallel to the $x$ axis 
with velocity $v$. The $[J_q]^{\gamma\delta}_{\gamma'\delta'}$ tensor in Eq. ~\ref{eq:RRIME-J} is then a product 
of two Lorentz tensors of the form of Eq.~\ref{eq:Lboost}. In the absence of other effects, the measured 
coherency becomes 

\begin{equation}
[V_{pq}]^{\alpha\beta\gamma\delta} = 
B^{\alpha\beta\gamma'\delta'}
[\Lambda_q]^\gamma_{\gamma'} [\Lambda_q]^\delta_{\delta'},
\end{equation}
or writing it out as an explicit sum for two particular elements of interest (and dropping the $pq$ indices):
\begin{eqnarray}
V^{0110} & = & \sum_{\gamma\delta}
B^{01\gamma\delta}
[\Lambda_q]^{0}_{\gamma} 
[\Lambda_q]^{1}_{\delta} \\
V^{0220} & = & \sum_{\gamma\delta}
B^{02\gamma\delta}
[\Lambda_q]^{0}_{\gamma} 
[\Lambda_q]^{2}_{\delta}.
\end{eqnarray}
Each sum above nominally contains 16 terms, but if we assume an unpolarized point source, then from Table~\ref{tab:BT} (looking up rows `01' and `02') we note that only four components of $B$ in each sum are non-zero. Combining this with Eq.~\ref{eq:Lboost}, we get:
\begin{equation}
V^{0110} = I(\gamma^2 - \beta^2\gamma^2) = I, 
\end{equation}
and
\begin{equation}
V^{0220} = I\gamma.
\end{equation} 
Doing the same sums for $V^{0120}$ and $V^{0210}$, we arrive at zero. This implies that our instrument will 
measure the Stokes parameters as
\begin{eqnarray}
I_\mathrm{meas} &=& \frac{V^{0110}+V^{0220}}{2} = I\frac{1+\gamma}{2} \\ 
Q_\mathrm{meas} &=& \frac{V^{0110}-V^{0220}}{2} = I\frac{1-\gamma}{2} \\
U_\mathrm{meas} &=& 0 \\
V_\mathrm{meas} &=& 0.
\end{eqnarray}

Since $\gamma>1$, we measure boost in the total power $I$, and negative apparent $Q$ (that is, linear polarization 
perpendicular to the direction of motion). The physical meaning of this instrumental $Q$ can be understood in
terms of the polarization aberration discussed by \citet{Carozzi:2009hf}: because the arriving plane wave is aberrated
in the frame of antenna $q$ (i.e. no longer appears to propagate along the $z$ axis, but along a slightly different 
direction), it is measured as polarized by dipoles that are parallel to $xy$.

Note that this formulation does not incorporate the Doppler shift observed by a moving antenna (we quietly assume that the 
correlator takes care of correcting for this when channelizing), or the problems of clock distribution to a relativistically
moving observing platform. This will have to be addressed in a future work.

In principle, the effects of Doppler boost and relativistic aberration can be described without invoking the full RRIME -- 
traditional Jones calculus still suffices. 
However the RRIME provides a unifying framework that allows these effects to be incorporated into a single compact interferometric 
measurement equation. This is similar to how the original RIME formulation of \citet{Hamaker:1996p5735} incorporated
polarimetric effects that were already described previously \citep{Sault-AT-Calibration} in a compact closed form.

An even more interesting use case arises when the incident EMF can no longer be described by a plane wave. 
The EMF tensor of Eq.~\ref{eq:F-general} then no longer reduces to Eq.~\ref{eq:F-planewave}, 
or in other words, the $\bmath{e}$ and $\bmath{h}$ components of the EM are no longer mutually redundant. Under Lorentz 
transformations, the $\bmath{e}$ and $\bmath{h}$ components then become intermixed -- an antenna measuring 
$\bmath{e}$ in the rest frame will also measure some contribution from $\bmath{h}$ in a moving frame. It is clear that 
in this case, Jones calculus (which only operates on $\bmath{e}$) can no longer apply, and a full RRIME must be invoked. 
Practical applications of this will have to wait for near-field space VLBI.

Conceptually, the latter example is very similar to the transmission matrix formalism. In a system where the voltage and 
current components intermix, the voltage-only Jones formalism no longer applies, and a formalism incorporating both 
components must be invoked.

\section{Conclusions}

The radio interferometer Measurement Equation provides a powerful
framework for describing a signal's journey from an astrophysical
source to the receiver of a radio telescope. Nevertheless, the Jones
calculus it employs is in general not sufficient to describe the components
within the analogue chain of a radio telescope. Within the telescope,
methods from microwave network theory, such as transmission matrices
and scattering parameters, are more appropriate. 

Similarly, the Jones formalism is not able to describe relativistic effects.  We have presented
two reformulations that account for mixed and magnetic field coherency
in a way that is not possible with the Jones formalism. These reformulations
extend the applicability of the RIME to allow it to correctly model analogue components,
and to describe relativistic effects within the RIME. 

\section*{Acknowledgments}

We would like to thank Stef Salvini, Ian Heywood and Mike Jones for 
their comments, Tobias Carozzi for his valuable insight, and the late Steve Rawlings for his advice 
in the seminal stages of this paper. O.~Smirnov's research is supported by the South African 
Research  Chairs Initiative of the Department of Science and Technology and National 
Research Foundation. This work has made use of LWA1 outrigger dipoles made available by the
LEDA project, funded by NSF under grants AST-1106054, AST-1106059, AST-1106045, and AST-
1105949.

\bibliographystyle{mn2e-lyx/mn2e}
\bibliography{references}

\bsp

\label{lastpage}
\end{document}